\documentclass{ws-procs961x669}            % default, citations in superscript

\usepackage{ifthen} % for conditional statements
\newboolean{articletitles}
\setboolean{articletitles}{true} % False removes titles in references
\newboolean{uprightparticles}
\setboolean{uprightparticles}{false} %True for upright particle symbols
\newboolean{inbibliography}
\setboolean{inbibliography}{false} %True once you enter the bibliography
\usepackage{xspace} 
\usepackage{upgreek}

\def\MagUp {\mbox{\em Mag\kern -0.05em Up}\xspace}

\ifthenelse{\boolean{uprightparticles}}%
{
 
 \def\Pgamma      {\ensuremath{\upgamma}\xspace}

 \def\Peta        {\ensuremath{\upeta}\xspace}

 \def\Pmu         {\ensuremath{\upmu}\xspace}                 
 \def\Pnu         {\ensuremath{\upnu}\xspace}                 
                  
 \def\Ppi         {\ensuremath{\uppi}\xspace}

 \def\Ptau        {\ensuremath{\uptau}\xspace}

 \def\Pchi        {\ensuremath{\upchi}\xspace}                 
 \def\Ppsi        {\ensuremath{\uppsi}\xspace}

 \def\PDelta      {\ensuremath{\Delta}\xspace}                 
 \def\PXi      {\ensuremath{\Xi}\xspace}                 
 \def\PLambda      {\ensuremath{\Lambda}\xspace}                 
 \def\PSigma      {\ensuremath{\Sigma}\xspace}                 
 \def\POmega      {\ensuremath{\Omega}\xspace}                 
 \def\PUpsilon      {\ensuremath{\Upsilon}\xspace}                 
 
 \def\PB      {\ensuremath{\mathrm{B}}\xspace}                 
                  
 \def\PD      {\ensuremath{\mathrm{D}}\xspace}

 \def\PH      {\ensuremath{\mathrm{H}}\xspace}                 
                  
 \def\PJ      {\ensuremath{\mathrm{J}}\xspace}                 
 \def\PK      {\ensuremath{\mathrm{K}}\xspace}

 \def\PW      {\ensuremath{\mathrm{W}}\xspace}

 \def\PZ      {\ensuremath{\mathrm{Z}}\xspace}                 
                  
 \def\Pb      {\ensuremath{\mathrm{b}}\xspace}                 
 \def\Pc      {\ensuremath{\mathrm{c}}\xspace}                 
 \def\Pd      {\ensuremath{\mathrm{d}}\xspace}                 
 \def\Pe      {\ensuremath{\mathrm{e}}\xspace}

 \def\Ph      {\ensuremath{\mathrm{h}}\xspace}                 
 \def\Pi      {\ensuremath{\mathrm{i}}\xspace}

 \def\Pn      {\ensuremath{\mathrm{n}}\xspace}                 
                  
 \def\Pp      {\ensuremath{\mathrm{p}}\xspace}                 
 \def\Pq      {\ensuremath{\mathrm{q}}\xspace}                 
                  
 \def\Ps      {\ensuremath{\mathrm{s}}\xspace}                 
 \def\Pt      {\ensuremath{\mathrm{t}}\xspace}                 
 \def\Pu      {\ensuremath{\mathrm{u}}\xspace}

}
{
 
 \def\Pgamma      {\ensuremath{\gamma}\xspace}

 \def\Peta        {\ensuremath{\eta}\xspace}

 \def\Pmu         {\ensuremath{\mu}\xspace}                 
 \def\Pnu         {\ensuremath{\nu}\xspace}                 
                  
 \def\Ppi         {\ensuremath{\pi}\xspace}

 \def\Ptau        {\ensuremath{\tau}\xspace}

 \def\Pchi        {\ensuremath{\chi}\xspace}                 
 \def\Ppsi        {\ensuremath{\psi}\xspace}                 
                  
 \mathchardef\PDelta="7101
 \mathchardef\PXi="7104
 \mathchardef\PLambda="7103
 \mathchardef\PSigma="7106
 \mathchardef\POmega="710A
 \mathchardef\PUpsilon="7107
                  
 \def\PB      {\ensuremath{B}\xspace}                 
                  
 \def\PD      {\ensuremath{D}\xspace}

 \def\PH      {\ensuremath{H}\xspace}                 
                  
 \def\PJ      {\ensuremath{J}\xspace}                 
 \def\PK      {\ensuremath{K}\xspace}

 \def\PW      {\ensuremath{W}\xspace}

 \def\PZ      {\ensuremath{Z}\xspace}                 
                  
 \def\Pb      {\ensuremath{b}\xspace}                 
 \def\Pc      {\ensuremath{c}\xspace}                 
 \def\Pd      {\ensuremath{d}\xspace}                 
 \def\Pe      {\ensuremath{e}\xspace}

 \def\Ph      {\ensuremath{h}\xspace}                 
 \def\Pi      {\ensuremath{i}\xspace}

 \def\Pn      {\ensuremath{n}\xspace}                 
                  
 \def\Pp      {\ensuremath{p}\xspace}                 
 \def\Pq      {\ensuremath{q}\xspace}                 
                  
 \def\Ps      {\ensuremath{s}\xspace}                 
 \def\Pt      {\ensuremath{t}\xspace}                 
 \def\Pu      {\ensuremath{u}\xspace}

}

\DeclareRobustCommand{\optbar}[1]{\shortstack{{\miniscule (\rule[.5ex]{1.25em}{.18mm})}
  \\ [-.7ex] $#1$}}

\def\en         {{\ensuremath{\Pe^-}}\xspace}   % electron negative (\em is taken)

\def\epem       {{\ensuremath{\Pe^+\Pe^-}}\xspace}

\def\mup        {{\ensuremath{\Pmu^+}}\xspace}
\def\mun        {{\ensuremath{\Pmu^-}}\xspace} % muon negative (\mum is taken)

\def\neu        {{\ensuremath{\Pnu}}\xspace}
\def\neub       {{\ensuremath{\overline{\Pnu}}}\xspace}
\def\neue       {{\ensuremath{\neu_e}}\xspace}
\def\neueb      {{\ensuremath{\neub_e}}\xspace}
\def\neum       {{\ensuremath{\neu_\mu}}\xspace}
\def\neumb      {{\ensuremath{\neub_\mu}}\xspace}
\def\neut       {{\ensuremath{\neu_\tau}}\xspace}
\def\neutb      {{\ensuremath{\neub_\tau}}\xspace}
\def\neul       {{\ensuremath{\neu_\ell}}\xspace}
\def\neulb      {{\ensuremath{\neub_\ell}}\xspace}
\def\g      {{\ensuremath{\Pgamma}}\xspace}

\def\dquark    {{\ensuremath{\Pd}}\xspace}

\def\squark    {{\ensuremath{\Ps}}\xspace}

\def\cquark    {{\ensuremath{\Pc}}\xspace}

\def\bquark    {{\ensuremath{\Pb}}\xspace}

\def\pion   {{\ensuremath{\Ppi}}\xspace}

\def\pip    {{\ensuremath{\pion^+}}\xspace}
\def\pim    {{\ensuremath{\pion^-}}\xspace}

\def\kaon    {{\ensuremath{\PK}}\xspace}
  \def\Kbar    {{\kern 0.2em\overline{\kern -0.2em \PK}{}}\xspace}

\def\KorKbar    {\kern 0.18em\optbar{\kern -0.18em K}{}\xspace}
\def\Kz      {{\ensuremath{\kaon^0}}\xspace}
\def\Kzb     {{\ensuremath{\Kbar{}^0}}\xspace}
\def\Kp      {{\ensuremath{\kaon^+}}\xspace}
\def\Km      {{\ensuremath{\kaon^-}}\xspace}

\def\KS      {{\ensuremath{\kaon^0_{\mathrm{ \scriptscriptstyle S}}}}\xspace}

\def\Kstarz  {{\ensuremath{\kaon^{*0}}}\xspace}
\def\Kstarzb {{\ensuremath{\Kbar{}^{*0}}}\xspace}
\def\Kstar   {{\ensuremath{\kaon^*}}\xspace}

  \def\Dbar    {{\kern 0.2em\overline{\kern -0.2em \PD}{}}\xspace}
\def\D       {{\ensuremath{\PD}}\xspace}

\def\DorDbar    {\kern 0.18em\optbar{\kern -0.18em D}{}\xspace}
\def\Dz      {{\ensuremath{\D^0}}\xspace}
\def\Dzb     {{\ensuremath{\Dbar{}^0}}\xspace}
\def\Dp      {{\ensuremath{\D^+}}\xspace}
\def\Dm      {{\ensuremath{\D^-}}\xspace}

\def\Ds      {{\ensuremath{\D^+_\squark}}\xspace}

\def\B       {{\ensuremath{\PB}}\xspace}
\def\Bbar    {{\ensuremath{\kern 0.18em\overline{\kern -0.18em \PB}{}}}\xspace}

\def\BorBbar    {\kern 0.18em\optbar{\kern -0.18em B}{}\xspace}

\def\Bu      {{\ensuremath{\B^+}}\xspace}
\def\Bub     {{\ensuremath{\B^-}}\xspace}
\def\Bp      {{\ensuremath{\Bu}}\xspace}

\def\Bd      {{\ensuremath{\B^0}}\xspace}
\def\Bs      {{\ensuremath{\B^0_\squark}}\xspace}

\def\Bdb     {{\ensuremath{\Bbar{}^0}}\xspace}
\def\Bc      {{\ensuremath{\B_\cquark^+}}\xspace}

\def\jpsi     {{\ensuremath{{\PJ\mskip -3mu/\mskip -2mu\Ppsi\mskip 2mu}}}\xspace}

  \def\Y#1S{\ensuremath{\PUpsilon{(#1S)}}\xspace}% no space before {...}!

\def\proton      {{\ensuremath{\Pp}}\xspace}

\def\neutron     {{\ensuremath{\Pn}}\xspace}

\def\Deltares    {{\ensuremath{\PDelta}}\xspace}

\def\Xires       {{\ensuremath{\PXi}}\xspace}

\def\Lz          {{\ensuremath{\PLambda}}\xspace}
\def\Lbar        {{\ensuremath{\kern 0.1em\overline{\kern -0.1em\PLambda}}}\xspace}
\def\LorLbar    {\kern 0.18em\optbar{\kern -0.18em \PLambda}{}\xspace}
\def\Lambdares   {{\ensuremath{\PLambda}}\xspace}

\def\Sigmares    {{\ensuremath{\PSigma}}\xspace}

\def\Omegares    {{\ensuremath{\POmega}}\xspace}

\def\Lb      {{\ensuremath{\Lz^0_\bquark}}\xspace}

\newcommand{\decay}[2]{\ensuremath{#1\!\to #2}\xspace}         % {\Pa}{\Pb \Pc}

\def\to                 {\ensuremath{\rightarrow}\xspace}

\def\CP                {{\ensuremath{C\!P}}\xspace}

\newcommand{\DGs}{{\ensuremath{\Delta\Gamma_{\squark}}}\xspace}

\newcommand{\phid}{{\ensuremath{\phi_{\dquark}}}\xspace}

\newcommand{\phis}{{\ensuremath{\phi_{\squark}}}\xspace}

\newcommand{\etag}{{\ensuremath{\varepsilon_{\mathrm{tag}}}}\xspace}

\def\AT#1     {\ensuremath{A_{\mathrm{T}}^{#1}}\xspace}           % 2

\def\C#1      {\ensuremath{\mathcal{C}_{#1}}\xspace}                       % 9
\def\Cp#1     {\ensuremath{\mathcal{C}_{#1}^{'}}\xspace}                    % 7
\def\Ceff#1   {\ensuremath{\mathcal{C}_{#1}^{\mathrm{(eff)}}}\xspace}        % 9  
\def\Cpeff#1  {\ensuremath{\mathcal{C}_{#1}^{'\mathrm{(eff)}}}\xspace}       % 7
\def\Ope#1    {\ensuremath{\mathcal{O}_{#1}}\xspace}                       % 2
\def\Opep#1   {\ensuremath{\mathcal{O}_{#1}^{'}}\xspace}                    % 7

\newcommand{\unit}[1]{\ensuremath{\mathrm{ \,#1}}\xspace}          % {kg}

\newcommand{\tev}{\ifthenelse{\boolean{inbibliography}}{\ensuremath{~T\kern -0.05em eV}\xspace}{\ensuremath{\mathrm{\,Te\kern -0.1em V}}}\xspace}
\newcommand{\gev}{\ensuremath{\mathrm{\,Ge\kern -0.1em V}}\xspace}
\newcommand{\mev}{\ensuremath{\mathrm{\,Me\kern -0.1em V}}\xspace}
\newcommand{\kev}{\ensuremath{\mathrm{\,ke\kern -0.1em V}}\xspace}
\newcommand{\ev}{\ensuremath{\mathrm{\,e\kern -0.1em V}}\xspace}
\newcommand{\gevc}{\ensuremath{{\mathrm{\,Ge\kern -0.1em V\!/}c}}\xspace}
\newcommand{\mevc}{\ensuremath{{\mathrm{\,Me\kern -0.1em V\!/}c}}\xspace}
\newcommand{\gevcc}{\ensuremath{{\mathrm{\,Ge\kern -0.1em V\!/}c^2}}\xspace}
\newcommand{\gevgevcccc}{\ensuremath{{\mathrm{\,Ge\kern -0.1em V^2\!/}c^4}}\xspace}
\newcommand{\mevcc}{\ensuremath{{\mathrm{\,Me\kern -0.1em V\!/}c^2}}\xspace}

\def\invfb   {\ensuremath{\mbox{\,fb}^{-1}}\xspace}

\def\invps{\ensuremath{{\mathrm{ \,ps^{-1}}}}\xspace}

\def\gsim{{~\raise.15em\hbox{$>$}\kern-.85em
          \lower.35em\hbox{$\sim$}~}\xspace}
\def\lsim{{~\raise.15em\hbox{$<$}\kern-.85em
          \lower.35em\hbox{$\sim$}~}\xspace}

\def\tell1  {TELL1\xspace}
\def\ukl1   {UKL1\xspace}

\usepackage{lineno}
\begin{document}
\title{$\CP$ violation in heavy-flavour hadrons}

\author{Greig A. Cowan, on behalf of the LHCb collaboration\footnote{28th International Symposium on Lepton Photon Interactions at High Energies, 
		7 - 12 Aug 2017\\
		Sun Yat-Sen University, Guangzhou, China\\
		email: g.cowan@ed.ac.uk}}

\address{University of Edinburgh, UK}

\begin{abstract}
Measurements of $\CP$-violating observables in $\B$ meson decays can be used to determine the angles of the Unitarity Triangle and hence probe for manifestations of New Physics beyond the Cabibbo-Kobayashi-Maskawa Standard Model paradigm.  Of particular interest are precise measurements of the angles $\gamma$ and $\beta$.   Also of great importance are studies of $\CP$-violation involving $\Bs$ mesons, in particular the phase $\phi_s$, which is a golden observable in flavour physics at the LHC.  Complementary to these studies is the continuing search for direct and indirect $\CP$-violation in the charm system, where the experimental precision is now at the $10^{-3}$ level.  I will present new and recent results in these topics, and in $\CP$-violation searches in baryon decays, with specific emphasis on the measurement programme at the LHC.
\end{abstract}

\keywords{Flavour physics; \CP violation; LHC}

\bodymatter

%\FullConference{28th International Symposium on Lepton Photon Interactions at High Energies\\
%		7 August - 12 August 2017\\
%		Sun Yat-Sen University, Guangzhou, China}

\section{Introduction}

The violation of $\CP$ symmetry, the combination of the discrete symmetries of charge-conjugation
(conjugation of all internal quantum numbers) and parity (reversing spatial coordinates),
is a necessary condition to generate the baryon asymmetry of the Universe~\cite{0038-5670-34-5-A08}.
However, the level of $\CP$ violation allowed within the quark sector of the SM is
many orders of magnitude too small~\cite{Gavela:1993ts,Huet:1994jb,Gavela:1994dt}
to explain astronomical observations, demanding experimental
searches for new sources of violation.
Heavy-quark hadrons provide an excellent laboratory to perform such searches as they allow the
exploration of high energy scales well beyond the direct reach of the LHC.
This approach has been successfully applied in the past with, for example, the observation of \Bd meson
mixing~\cite{Albajar:1986it,Prentice:1987ap} leading to the first estimates of the top quark mass before
it was directly discovered.
The large heavy-quark production cross-sections at the
LHC~\cite{LHCb-PAPER-2016-031,LHCb-PAPER-2015-041} lead to large
samples of exclusively reconstructed $b$ and $c$ hadron decays that have been used to make
precision measurements of \CP violating observables. These proceedings  summarise the latest of
these measurements, focussing on those using 3\invfb of data collected by the
LHCb experiment~\cite{Alves:2008zz} in $pp$ collisions
at the LHC during 2011 and 2012, unless otherwise stated.
Recent comprehensive reviews can be found in Refs.~\cite{Artuso:2015swg,Gershon:2016fda} and
references therein.

\section{\CP violation in the Standard Model}

The only source of \CP violation within the SM is due to the non-zero value of the phase in the 
Cabibbo-Kobayashi-Maskawa (CKM) matrix, to which all \CP-violating observables are related.
The unitarity of the matrix leads to relations between elements (e.g., 
$V_{ud}V^*_{ub} + V_{cd}V^*_{cb} + V_{td}V^*_{tb} = 0$), which are convenient to visualise as a 
triangle in the complex plane.  For the triangle ({\it the} Unitarity Triangle) 
where all sides are of a similar magnitude, the angles are defined as 
\begin{center}
$\alpha    \equiv \arg\left[-\frac{V_{td}V^*_{tb}}{V_{ud}V^*_{ub}}\right]$,\,\;
$\beta      \equiv \arg\left[-\frac{V_{cd}V_{cb}^*}{V_{td}V_{tb}^*}\right]$,\,\;
$\gamma \equiv \arg\left[-\frac{V_{ud}V^*_{ub}}{V_{cd}V^*_{cb}}\right]$.
\end{center}

\begin{figure}[h]
\begin{center}
\includegraphics[width=0.45\linewidth]{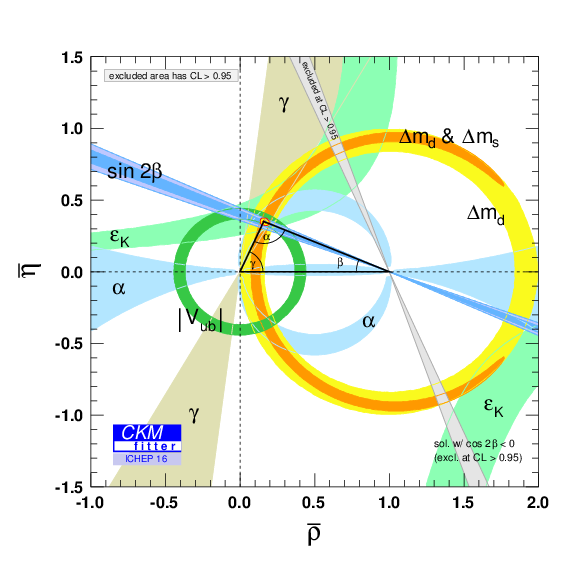}
\end{center}
\caption{\label{fig:CKM}\small Global fit~\cite{Charles:2015gya} to the CKM matrix parameters, showing
consistency between measurements when interpreted in terms of SM quark transitions.}
\end{figure}

These angles can be measured using a variety of different \CP violating observables covering both
tree-level quark transitions, where the impact of New Physics (NP) contributions is expected to be
small~\cite{Brod:2013sga},
and loop-level transitions, which are sensitive to new higher-mass particles. One of the goals of
studying the heavy-quark sector is to compare measurements of these quantities to check for the 
overall consistency of the CKM mechanism. 
Figure~\ref{fig:CKM} shows the latest global fit of the CKM matrix parameters to experimental 
measurements and Lattice QCD calculations~\cite{Bazavov:2016nty,Aoki:2016frl}
showing that the SM is working well. However, there 
is still room for NP contributions at the level of $\sim 10\%$~\cite{Charles:2015gya,Bona:2006ah},
implying that a new set of precision measurements is required.

In the quark sector, the neutral mesons ($P^0$) can oscillate into their antiparticles
($\overline{P}^0$), resulting in the 
physical states ($P^0_{\rm H, L}$) being admixtures of the flavour eigenstates:
$P^0_{\rm H} = p P^0 + q \overline{P}^0$ and
$P^0_{\rm L} = p P^0 - q \overline{P}^0$,
where $p$ and $q$ are complex coefficients ($|p|^2+|q|^2 = 1$).
The physical states have well defined masses and lifetimes, and the parameters
$\Delta m = m_{\rm H} - m_{\rm L}$, $\Delta\Gamma = \Gamma_{\rm L} - \Gamma_{\rm H}$
and $\Gamma = (\Gamma_{\rm L} + \Gamma_{\rm H})/2$ control the decay-time-dependent
decay rates of the mesons. In particular, $\Delta m$ controls the oscillation frequency.

\CP violation depends on the quantity $\lambda_{f} = \frac{q}{p}\frac{\overline{A}_{f}}{A_{f}}$.
Here, $f$ is the final state that the $P$ hadron decays to with amplitude $A_{f}$
and the $\overline{P}$ hadron decays to with amplitude $\overline{A}_{f}$. Three types of
\CP violation are allowed: in neutral meson mixing ($|q/p| \neq 1$); 
in the interference between neutral meson
mixing and decay \mbox{($\arg(\lambda_{f}) \neq 0$)} and in hadron decay
($|\overline{A}_{f}/A_{f}| \neq 1$).
Only \CP-violation in decay is allowed for charged mesons and baryons.

%\subsection{Typical analysis ingredients}
%
%Excellent decay-time resolution~\cite{LHCb-PAPER-2014-059}
%Modelling decay-time efficiency.
%Production + detection asymmetries.
%Tagging of meson flavour @ production~\cite{LHCb-PAPER-2016-039}.
%Typical tagging power
%~ 4\%   LHCb (J/? modes) 
%~ 8\%   LHCb (open-charm modes)
%~1.5\%  ATLAS/CMS
%~ 30\% B-factories

\section{\CP violation in \B meson mixing}

%The typical method to look for \CP violation in \B meson mixing is to
%use semileptonic $B^0_{(s)}$ decays. Since these are dominated by tree-level quark transitions
%there should be no \CP violation in decay, providing a clean probe of \CP in mixing.

Semileptonic $B^0_{(s)}$ decays are dominated by tree-level quark transitions, implying that there should be no \CP violation in decay, and therefore provide a clean system to search for \CP in mixing.
The so-called semileptonic (or flavour-specific) asymmetry is defined as
\mbox{$A_{\rm sl} = \frac{\Gamma(\overline{B}^0 \to B^0 \to f) - \Gamma(B^0 \to \overline{B}^0  \to \overline{f})}{\Gamma(\overline{B}^0 \to B^0 \to f) + \Gamma(B^0 \to \overline{B}^0  \to \overline{f})} \approx \frac{\Delta\Gamma}{\Delta m}\tan\phi_{M}$}, where $\phi_M$ is the mixing phase from the $B^0_{(s)}$ mixing matrix.
These asymmetries are predicted to be very small in the SM, at the level of $10^{-4}$ or less~\cite{Artuso:2015swg}, 
and therefore any measurement of a significantly non-zero effect would be a clear sign of beyond-the-SM physics.

Experimentally, the quantity measured is the untagged decay-time, $t$, dependent
 charge asymmetry between semileptonic $B^0_{(s)}$ decays
with a positive or negatively charged muon, defined as 
\mbox{$A_{\rm meas}(t) = \frac{N(D^-\mu^+\nu,t) - N(D^+\mu^-\nu,t)}{N(D^-\mu^+\nu,t) + N(D^+\mu^-\nu,t)}$} \mbox{$\approx A_{\rm D} + \frac{A_{\rm sl}}{2} + \left(A_{\rm P} - \frac{A_{\rm sl}}{2}\right)\cos(\Delta m t)$}. This is sensitive to $A_{\rm sl}$ along with other production, 
$A_{\rm P}$, and particle detection, $A_{\rm D}$, asymmetries.
In the case of the measurements from the LHCb
collaboration~\cite{LHCb-PAPER-2016-013,LHCb-PAPER-2014-053} these asymmetries
can be controlled to high precision using data calibration samples and by reversing the LHCb dipole magnet,
thereby allowing a precision measurement of the \CP asymmetries in both the \Bd and \Bs systems.\footnote{ 
For the \Bs system, the time-integrated
rate can be measured as the fast \Bs oscillations wash out the production asymmetry.}
Figure~\ref{fig:asl}a
shows the current experimental situation for
\CP violation in \B meson mixing. The global average values are $A_{\rm sl}^d = (-0.21 \pm 0.17)\%$ and
$A_{\rm sl}^s = (-0.06 \pm 0.28)\%$~\cite{Amhis:2016xyh}, consistent with SM expectations. 

The left-most green ellipse in Figure~\ref{fig:asl}a corresponds to the dimuon asymmetry measured by the 
D0 collaboration\cite{Abazov:2013uma}, which is a measurement of a linear combination of 
$A_{\rm sl}^d$ and $A_{\rm sl}^s$. Although it is inconsistent with SM expectations at $\sim3\sigma$
it has been proposed~\cite{Borissov:2013wwa,Amhis:2015xea}
that there may be additional contributions from a non-zero value of $\Delta\Gamma_d/\Gamma_d$, 
which is expected to be very small in the SM. 
The most precise measurement of this quantity has recently been made by the ATLAS
collaboration~\cite{Aaboud:2016bro} (Figure~\ref{fig:asl}b), obtaining 
$\Delta\Gamma_d/\Gamma_d = (-0.1 \pm 1.1 \pm 0.9) \times 10^{-2}$.
An update of the 1\invfb measurement~\cite{LHCb-PAPER-2013-065}
of $\Delta\Gamma_d/\Gamma_d$ from the LHCb collaboration is eagerly anticipated.
%$A_{\CP} = C_d A_{\rm SL}^d + C_s A_{\rm SL}^s + C_{\Delta\Gamma_d} \Delta\Gamma_d/\Gamma_d$.

\def\figsubcap#1{\par\noindent\centering\footnotesize(#1)}

\begin{figure}[t]%
\begin{center}
  \parbox{2.1in}{\includegraphics[width=2in]{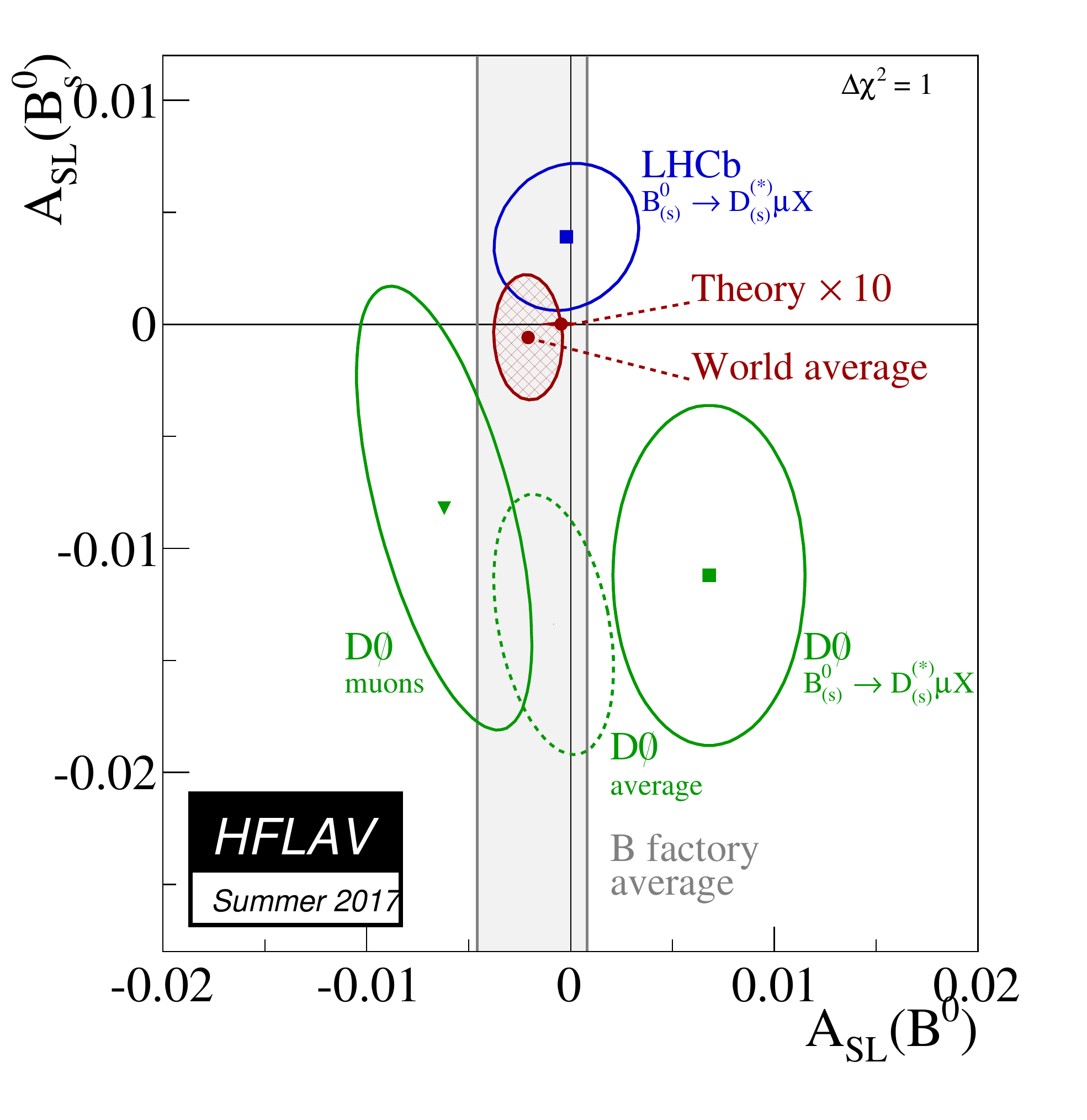}\figsubcap{a}}
  \hspace*{4pt}
  \parbox{2.1in}{\includegraphics[width=2.5in]{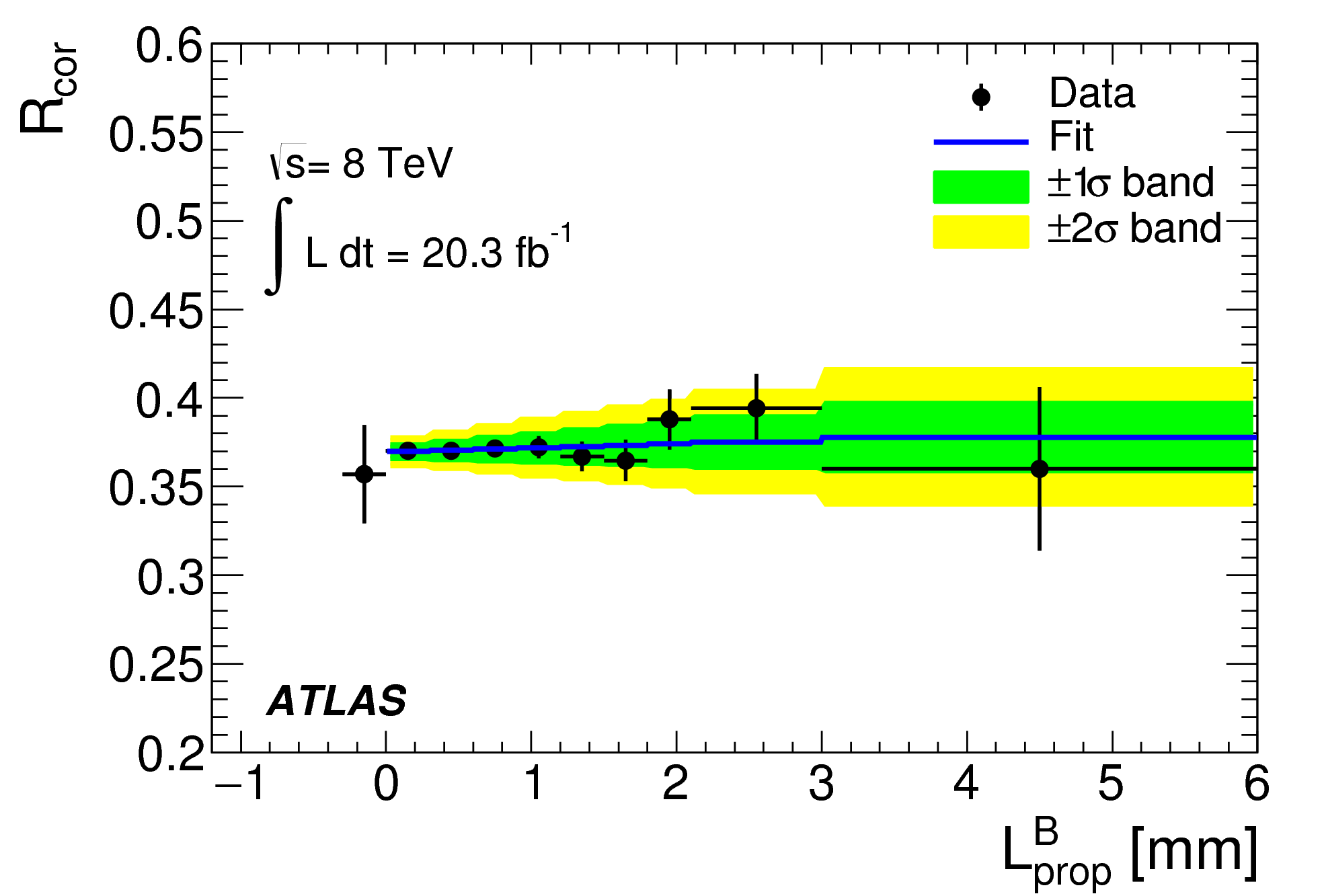}\figsubcap{b}}
\caption{\label{fig:asl}\small (a) HFLAV combination~\cite{Amhis:2016xyh} of $A_{\rm sl}^d$ and 
$A_{\rm sl}^s$ from several experiments compared to the theory ($\times 10$)
predictions~\cite{Artuso:2015swg}. (b) Efficiency-corrected ratio of the observed decay-length distributions
for $\Bd\to\jpsi\Kstar$ and \mbox{$\Bd\to\jpsi\KS$} decays~\cite{Aaboud:2016bro}.}
\end{center}
\end{figure}

\section{\CP violation in the interference of \B meson mixing/decay}

In the case where the $B^0_{(s)}$ or  $\overline{B}^0_{(s)}$ mesons decay to the same final state, $f$, the decay-time-dependent
\CP asymmetry is given by

\begin{equation}
\frac{\Gamma_{\overline{B}^0 \to f}(t) - \Gamma_{B^0 \to f}(t)}{\Gamma_{\overline{B}^0 \to f}(t) + \Gamma_{B^0 \to f}(t)} = \frac{S_f\sin(\Delta mt) - C_f\cos(\Delta mt)}{\cosh(\Delta\Gamma t/2) + A_f^{\Delta\Gamma}\sinh(\Delta\Gamma t/2)},
\label{eqn:asymm}
\end{equation}
where $|S_f|^2 + |C_f|^2 + |A_f^{\Delta\Gamma}|^2 = 1$ by definition. 

\subsection{The \Bd system}

In the \Bd system, $\Delta\Gamma_d \approx 0$ and only the numerator of Eq.~\ref{eqn:asymm}
needs to be considered. 
The canonical decay mode used by the B-factories to measure this asymmetry is $\Bd\to\jpsi\KS$,
which proceeds predominately via a tree-level $b\to c\overline{c}s$ transition. 
In the case where the sub-dominant penguin diagrams can be
neglected~\cite{Faller:2008zc,Jung:2012mp,DeBruyn:2014oga,Frings:2015eva}, \mbox{$S_{\jpsi\KS} \approx \sin2\beta$}.

\begin{figure}[t]%
\begin{center}
  \parbox{2.1in}{\includegraphics[width=1.05\linewidth]{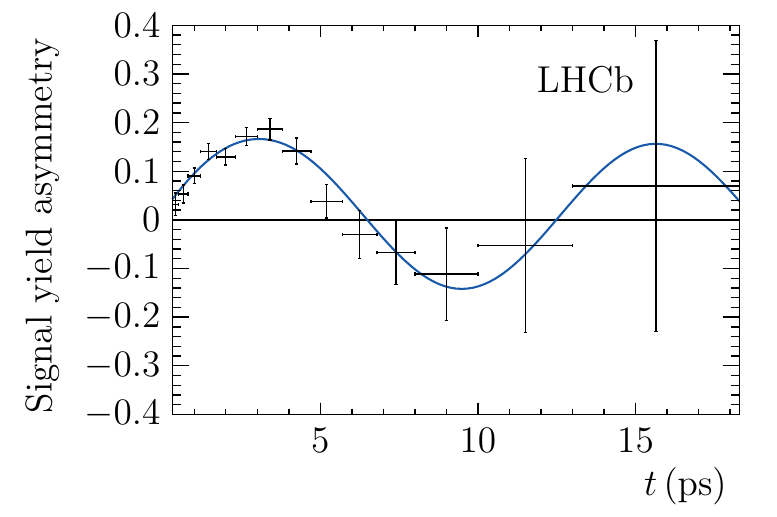}\figsubcap{a}}
  \hspace*{4pt}
  \parbox{2.1in}{\includegraphics[width=\linewidth,trim={8cm 0 0 0},clip]{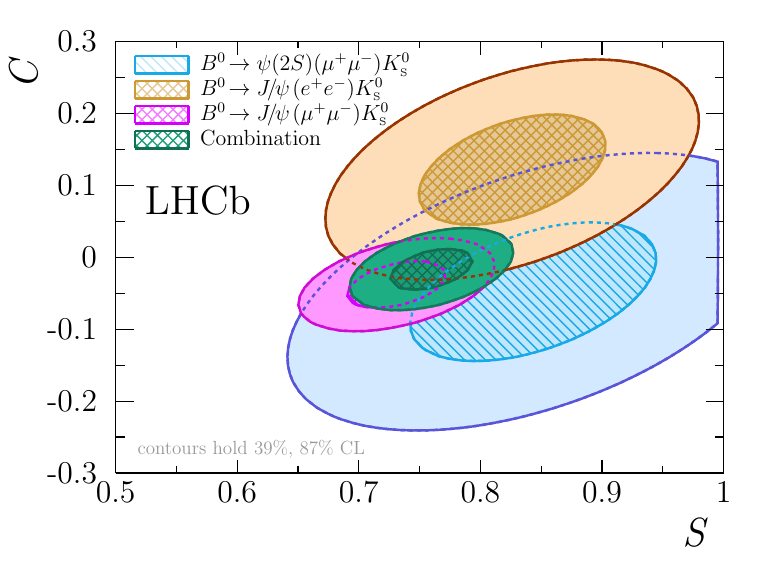}\figsubcap{b}}
\caption{\label{fig:sin2beta}\small  (a) \CP asymmetry as a function of decay time for $\Bd\to \jpsi\KS$
decays~\cite{LHCb-PAPER-2015-004}, showing a clear oscillation. (b) 
LHCb combination of \CP violation parameters as measured using
$\Bd\to \jpsi(\mu^+\mu^-)\KS$~\cite{LHCb-PAPER-2015-004},
$\Bd\to \jpsi(e^+e^-)\KS$ and  $\Bd\to\psi(2S)(\mu^+\mu^-)\KS$~\cite{LHCb-PAPER-2017-029}
decays.}
\end{center}
\end{figure}

The LHCb collaboration has recently used its Run 1 data to measure $S_f$ 
and $C_f$ in $\Bd\to\jpsi(\mu^+\mu^-)\KS$~\cite{LHCb-PAPER-2015-004},
$\Bd\to\jpsi(e^+e^-)\KS$ and $\Bd\to\psi(2S)(\mu^+\mu^-)\KS$ decays~\cite{LHCb-PAPER-2017-029}
using a flavour-tagged~\cite{LHCb-PAPER-2016-039} decay-time-dependent analysis. The 
asymmetry for $\Bd\to\jpsi(\mu^+\mu^-)\KS$ decays can be seen in Figure~\ref{fig:sin2beta}a.
The individual measurements and their combination are  shown in Figure~\ref{fig:sin2beta}b, where
the systematic uncertainty is dominated by background tagging asymmetry. The LHCb-averaged values
are $S_{[c\overline{c}]\KS} = 0.760 \pm 0.034$ and $C_{[c\overline{c}]\KS} = -0.017 \pm 0.029$.
Together, these measurements reduce the tension between the world average value for $\sin2\beta$
and the indirect determination from global fits~\cite{Charles:2015gya,Bona:2006ah}.
The consistency between the results using the electron and muon channels for charmonium reconstruction
also help to build confidence in the electron reconstruction performance of LHCb, which is 
particularly relevant when viewed through the prism of recent anomalies 
in $b\to s\ell^+\ell^-$ transitions~\cite{monica}.

\begin{figure}[t]
\begin{center}
\includegraphics[width=0.4\linewidth]{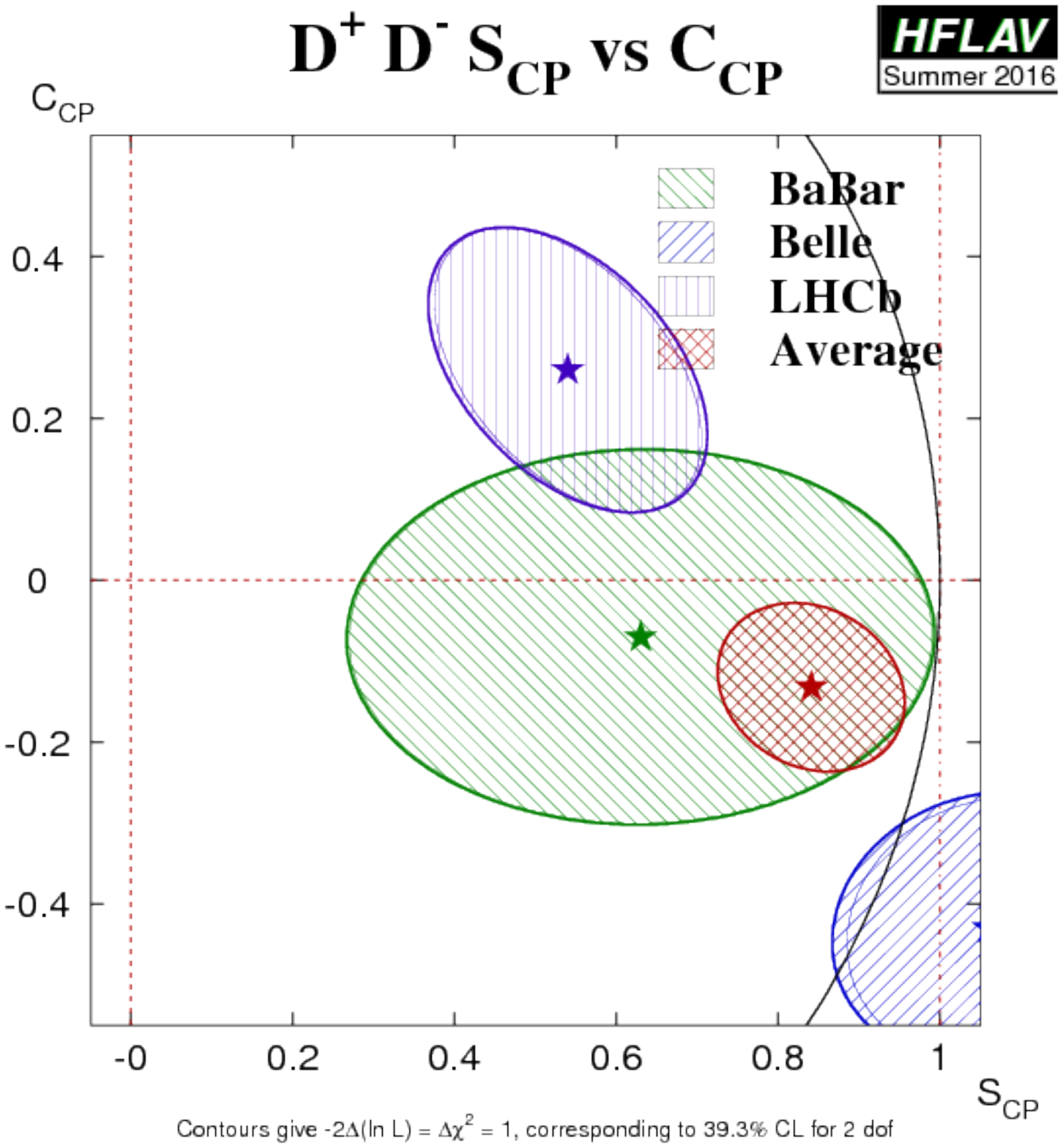}
\caption{\label{fig:B2DD}\small  HFLAV average of \CP parameters in $\Bd\to\Dp\Dm$ decays~\cite{Amhis:2016xyh}.}
\end{center}
\end{figure}

Decays such as $\Bd\to\Dp\Dm$ are governed by $b \to c\overline{c}d$ quark transitions and
therefore measurements of the decay-time-dependent asymmetry gives complimentary
information about $\sin2\beta$ that can be used to constrain the size of potential penguin
contributions to the decay~\cite{Faller:2008zc,Jung:2012mp,DeBruyn:2014oga,Frings:2015eva}.
Figure~\ref{fig:B2DD} summarises the
current situation with these measurements from the BaBar, Belle and
LHCb~\cite{LHCb-PAPER-2016-037} collaborations. The Belle result is outside of the physical region
($S_{DD}^2 + C_{DD}^2 < 1$), which may have been an indication of large hadronic effects.
However, the latest LHCb measurement
shows that these terms are small and consistent with zero, 
with the phase shift induced by the penguin diagrams measured to be
$\Delta\phi = -0.16^{+0.19}_{-0.21}$ rad.

\subsection{The \Bs system}

Decay-time-dependent \CP asymmetries in the \Bs system using $b\to c\overline{c}s$ 
transitions are sensitive to the CKM phase
$\beta_s     \equiv \arg\left[-\frac{V_{ts}V_{tb}^*}{V_{cs}V_{cb}^*}\right]$.
Typically measurements are made of the experimentally observable phase
$\phi_s$, which is equal to $-2\beta_s$ if the penguin contributions to the
decay can be neglected.
Global fits give a precise Standard Model prediction
for $\phi_s$ of $-36.5\pm 1.3$ mrad~\cite{Charles:2015gya}.
Deviations from this value would be a clear sign for NP,
strongly motivating the need for more precise experimental measurements.

The golden mode for measuring \phis\ is using a flavour-tagged
decay-time-dependent angular analysis of the $\Bs\to \jpsi(\mu^+\mu^-)\phi(\Kp\Km)$
decay. This channel has a high branching fraction
and the presence of two muons in the final state leads to a high trigger
efficiency at hadron colliders. An angular analysis is necessary to disentangle the interfering
\CP-odd and \CP-even components in the final state, which arise due to the
relative angular momentum between the two vector resonances.
In addition, there is a small ($\sim 2\%$) \CP-odd $\Kp\Km$ S-wave
contribution that must be accounted for.

\begin{figure}[t]%
\begin{center}
  \parbox{2.1in}{\includegraphics[width=\linewidth]{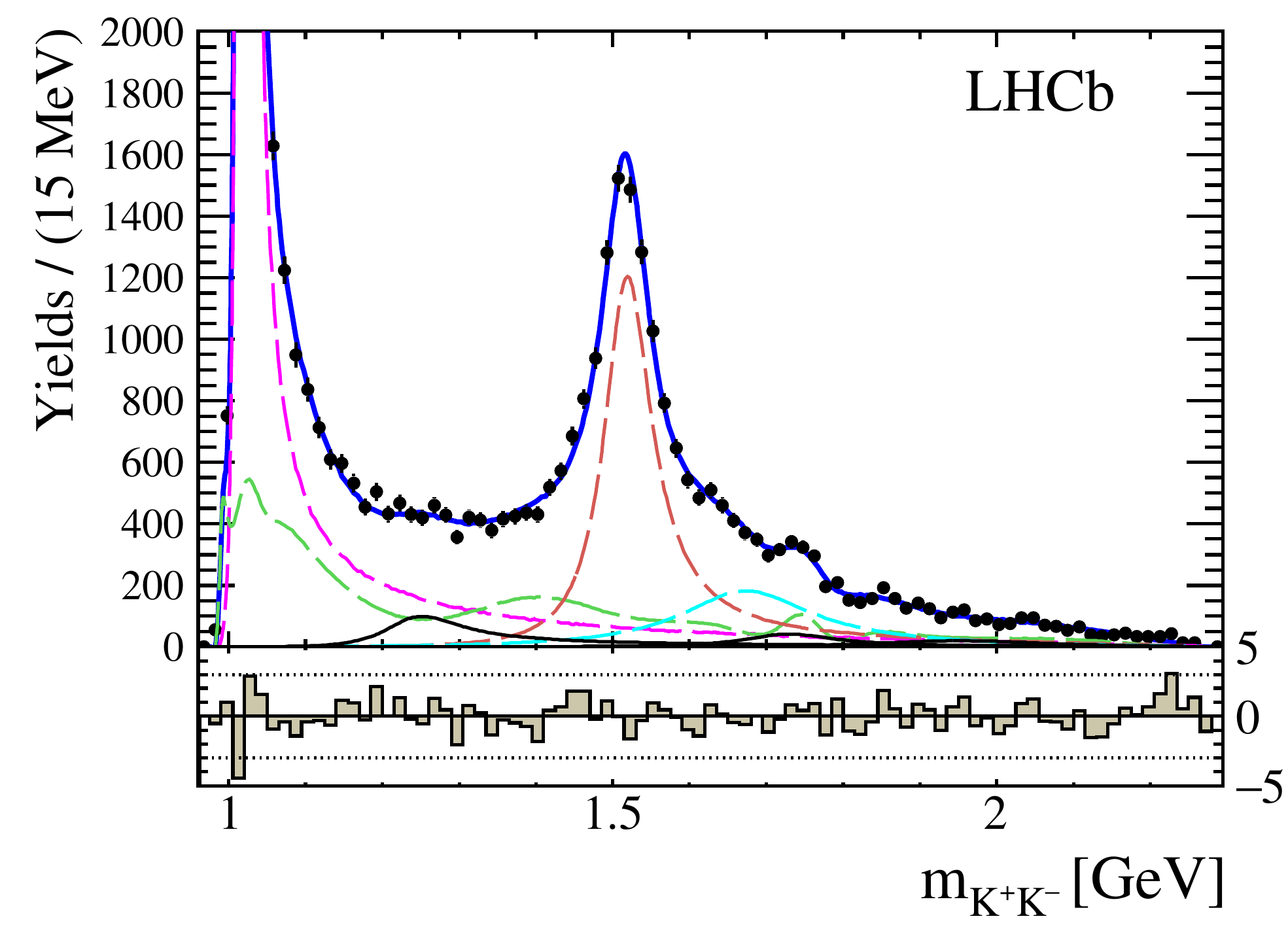}\figsubcap{a}}
  \hspace*{4pt}
  \parbox{2.1in}{\includegraphics[width=\linewidth]{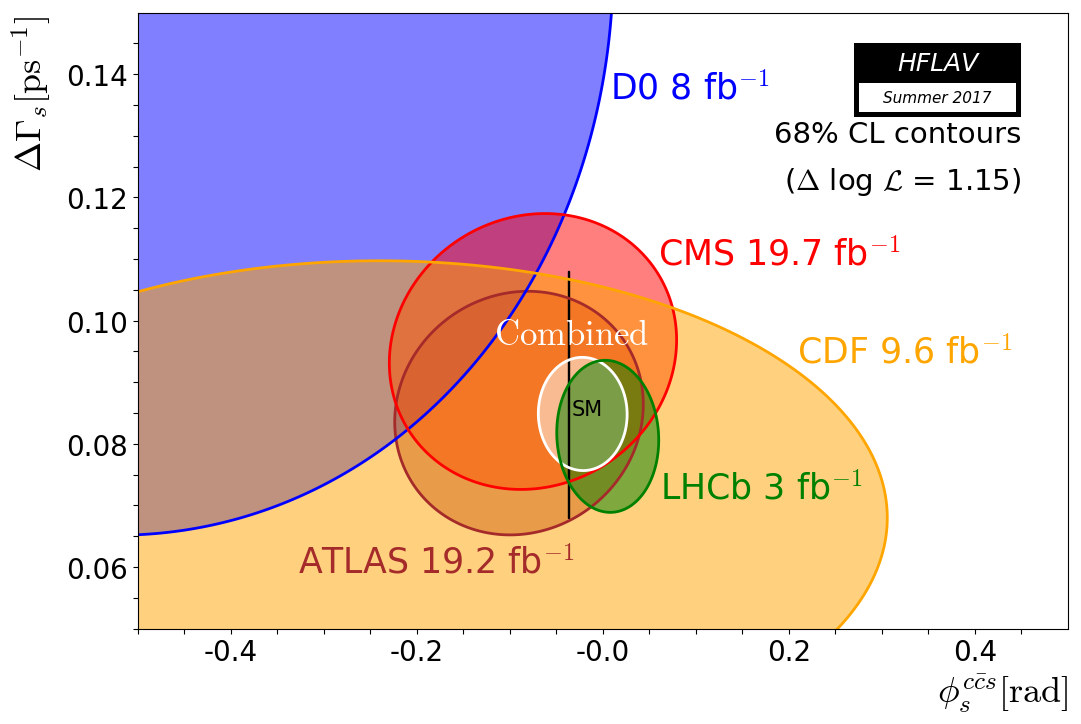}\figsubcap{b}}
\caption{\label{fig:phis}\small (a)  Distribution of $m_{\Kp\Km}$ from $\Bs\to\jpsi\Kp\Km$ decays. The
total fit function is overlaid in blue, while the
$\phi(1020)$, $f^\prime_2(1525)$ and $\Kp\Km$ S-wave contributions are shown by the long-dashed pink,
 brown and green lines, respectively. (b) 
HFLAV combination~\cite{Amhis:2016xyh} (white contour) of \phis\ and $\Delta\Gamma_s$ from several
experiments (coloured contours) as discussed in the text.}
\end{center}
\end{figure}

The CDF~\cite{Aaltonen:2012ie}, D0~\cite{Abazov:2011ry},
ATLAS~\cite{Aad:2016tdj}, CMS~\cite{Khachatryan:2015nza} and
LHCb~\cite{LHCb-PAPER-2014-059} collaborations have all measured $\phi_s$
(in addition to other mixing-related parameters of the \Bs system)
using the $\Bs\to \jpsi\phi$ decay. Additional information can be obtained by utilising the
region of the $\Kp\Km$ invariant mass spectrum above the $\phi(1020)$
meson, where higher spin $\Kp\Km$ resonances are expected to contribute.
Such a flavour-tagged decay-time-dependent amplitude analysis has just been performed by the LHCb collaboration~\cite{LHCb-PAPER-2017-008}, which finds the dominant component of the high-mass
spectrum comes from the $f^\prime_2(1525)$ meson (Figure~\ref{fig:phis})
and measures  $\phi_s = 119 \pm  107  \pm  34$ mrad.
The LHCb detector has excellent time resolution ($\sim 45$ fs)
and tagging power ($\sim 4\%$), both of which are crucial to the measurement.
Combining the LHCb results from
$\Bs\to \jpsi\phi$ (low mass), $\Bs\to \jpsi\Kp\Km$ (high mass) and
$\Bs\to \jpsi\pi^+\pi^-$ decays~\cite{LHCb-PAPER-2014-019} gives $\phis = 1  \pm  37$ mrad.

The global combination of \phis\ and $\Delta\Gamma_s$ using the measurements referenced above
in addition to $\Bs\to\psi(2S)\phi$~\cite{LHCb-PAPER-2016-027} and
$\Bs\to\Ds\Ds$\cite{LHCb-PAPER-2014-051} decays gives average values of 
$\DGs    =  0.090   \pm  0.005 \invps$ and $\phi_s  =  -21  \pm  31$~mrad.
The combination is dominated by the statistical uncertainty from the  
LHCb $\Bs\to\jpsi\phi$ result and are consistent with the SM
predictions~\cite{Charles:2015gya,Artuso:2015swg}. However, there remains
space for new physics contributions at  $O(10\%)$ and as the experimental 
precision improves it is essential that there is good control over hadronic
effects~\cite{Faller:2008gt,Bhattacharya:2012ph}
that could mimic the signature of beyond-the-SM physics.

A related \CP-violating phase, $\phi_s^{s\overline{s}s}$, can be
measured by applying similar analysis methods to \Bs meson decays that
occur via $b \to s\overline{s}s$ transitions. The LHCb collaboration has 
performed such an analysis using $\Bs\to \phi\phi$~\cite{LHCb-PAPER-2014-026},
measuring $\phis = -0.17 \pm 0.15 \pm 0.03$ rad, which is consistent with the SM
predictions, all of which are very close to zero~\cite{Beneke:2006hg,Bartsch:2008ps,Cheng:2009mu}.
Updated measurements of \CP-violating parameters in charmless $\Bs\to\Kp\Km$ decays
have also recently been reported by the LHCb collaboration~\cite{LHCb-CONF-2016-018},
including a first measurement of $A^{\Delta\Gamma}_{\Kp\Km}$.

\section{\CP violation in $b$ hadron decay}

\subsection{The CKM angle $\gamma$}

The CKM angle $\gamma$ is the only \CP violating parameter that can be measured from
tree-level decays\footnote{There are some caveats.} and the uncertainty on its theoretical
prediction is constrained to be $<O(10^{-7})$~\cite{Brod:2013sga}. Together these make 
measurements of $\gamma$ a ``standard candle'' within the SM with which other loop-level
determinations can be compared in order to look for the effects of new \CP violating
contributions. The canonical technique to measure $\gamma$ is to exploit the interference between 
the different decay paths in $B\to DK$ decays.
Depending on the final state of the $D^0$ meson there are different analysis
methods (e.g., GLW~\cite{Gronau:1990ra,Gronau:1991dp}, ADS\cite{Atwood:1996ci,Atwood:2000ck}
and GGSZ~\cite{Giri:2003ty}) that vary in their sensitivity to $\gamma$ and the other hadronic parameters
that describe the strong dynamics of the \B and \D meson decays. Unlike the measurement of
$\beta_{(s)}$, there is no dominant channel in which to measure $\gamma$ and a combination of several
modes is required to achieve the maximal sensitivity.

\begin{figure}[t]
\centering
\includegraphics[width=0.75\linewidth]{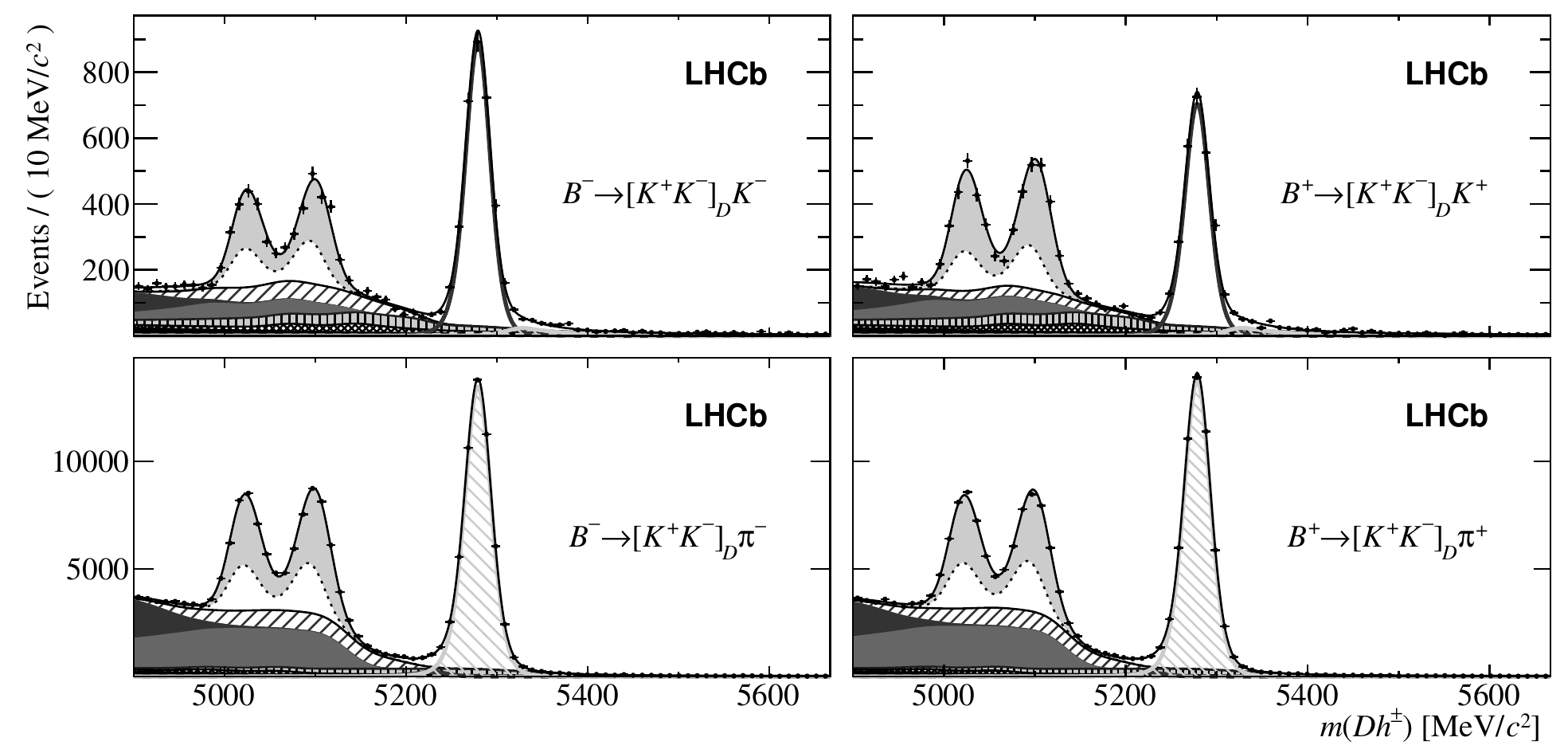}
\caption{\label{fig:B2DK}\small Distribution of invariant mass of the $Dh^\pm$ system for (left) $B^-$ and (right) 
$B^+$ decays. The bachelor hadron is a (top) kaon and (bottom) pion. In the case of the  $\B^\pm \to [\Kp\Km]_DK^\pm$ decays a clear asymmetry is visible between the distributions. The structures at lower
masses is due to partially reconstructed $D^{*0}\to D^0\gamma$ and $D^{*0}\to D^0\pi^0$ decays where the $\gamma$
or $\pi^0$ particle is missed.}
\end{figure}

A new result~\cite{LHCb-PAPER-2017-021} from the LHCb collaboration is an update, using Run 2 data,
of the measurements of the \CP-violating observables in 
$B^\pm \to D^{(*)0}K^{\pm}$ and $B^\pm \to D^{(*)0}\pi^{\pm}$
decays using the GLW method. Figure~\ref{fig:B2DK} shows the invariant mass distribution of the
$D^{0}h$ system. In the case of $B^\pm \to D^{0}K^{\pm}$ decays a clear asymmetry is visible between the 
oppositely-charged modes. After controlling for small detector and production asymmetries using the 
Cabibbo-favoured  $B^\pm \to [K^\pm\pi^\mp]_D\pi^{\pm}$ mode, the \CP asymmetry is measured to be
$A_K^{KK} = +0.126 \pm 0.014 \pm 0.002$, where the first uncertainty is statistical and the second 
systematic.

For the first time the collaboration has also used a partial reconstruction technique to measure
the \CP observables using the modes with excited $D^{*0}\to\Dz\gamma$ and $D^{*0}\to\Dz\pi^0$ decays where
the  photon or $\pi^0$ is not reconstructed.  This approach avoids the efficiency penalty that would
need to be paid to fully reconstruct these channels. 
The partially reconstructed decays correspond to the
structures at lower mass in Figure~\ref{fig:B2DK}, where the different shapes for the
$D^{*0}\to\Dz\gamma$ and $D^{*0}\to\Dz\pi^0$ contributions allow the decay rates and \CP asymmetries
to be measure separately for each.
The \CP asymmetry in the 
$B^\pm \to (D^{*0}\to\Dz\pi^0)K^{\pm}$ channel is measured as 
$A_K^{\CP, \pi^0} = -0.151 \pm 0.033 \pm 0.011$, which is different from zero at $4.3\sigma$. 
The corresponding asymmetry for the $\gamma$ mode is 
$A_K^{\CP, \gamma} = +0.276 \pm 0.094 \pm 0.047$.

\begin{figure}[t]
\centering
\includegraphics[width=0.8\linewidth]{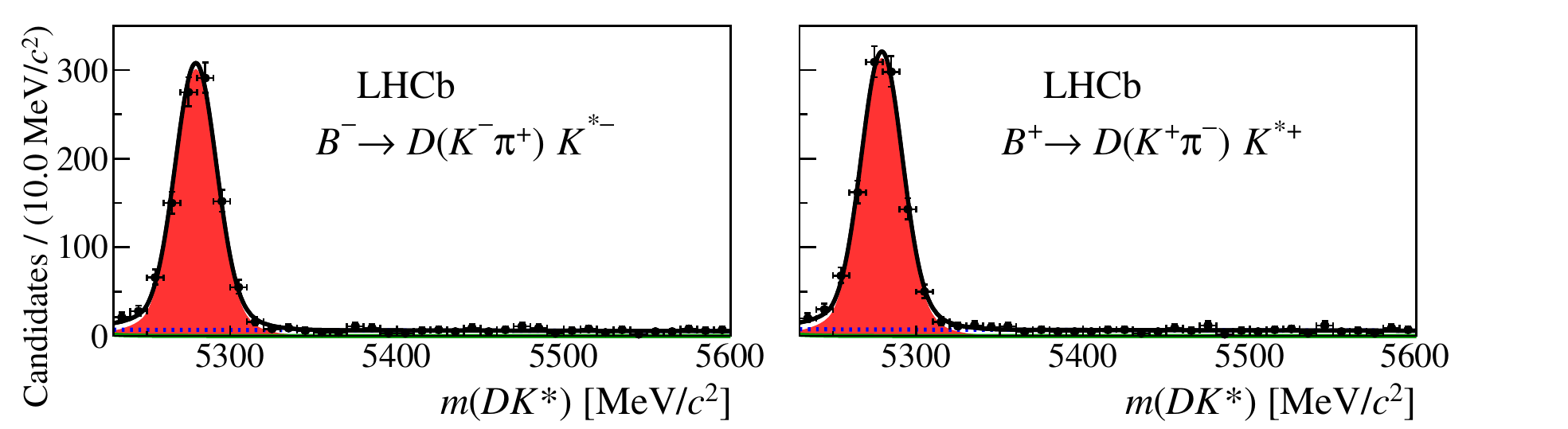}
\includegraphics[width=0.8\linewidth]{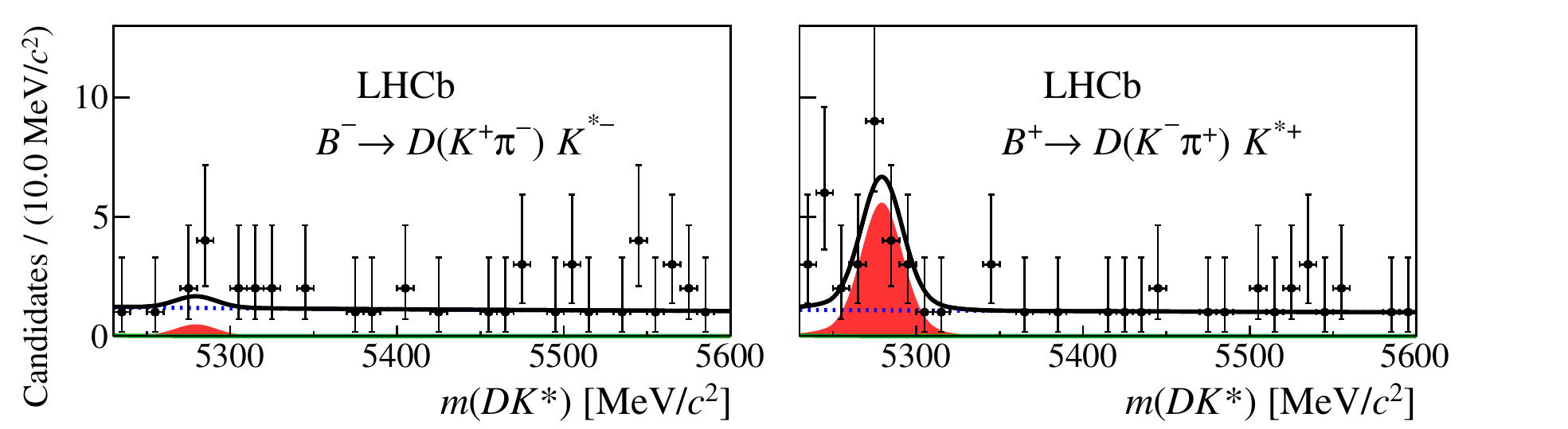}
\caption{\label{fig:B2DKst}\small Distribution of invariant mass of the $DK^*$ system for (left) $B^-$ and (right) 
$B^+$ decays. Both the (top) Cabibbo-favoured and (bottom) ADS $D^0\to K\pi$ decay modes are visible.}
\end{figure}

Another new result  is the measurement of
the \CP-violating observables in \mbox{$B^\pm \to DK^{\ast \pm}$} decays, with $K^{\ast \pm} \to \KS\pi^\pm$.
This updates Ref.~\cite{LHCb-CONF-2016-014}, using two- and four-body $D$ meson final
states~\cite{LHCb-PAPER-2017-030} in addition to Run 2 data. 
The branching ratio of the \mbox{$B^\pm \to DK^{\ast \pm}$} decay is of similar magnitude to 
\mbox{$B^\pm \to DK^{\pm}$} (described above), but the overall event yield in LHCb
is lower due to the efficiency for reconstructing $\KS$ mesons.
Figure~\ref{fig:B2DKst} shows the invariant mass distributions of the $DK^*$
systems for the two-body Cabibbo-favoured and ADS \Dz decay modes. The decays are isolated
with high signal purity and this result corresponds to the first $4.2\sigma$ evidence of the ADS mode,
with the rate measured to be \mbox{$R_{K\pi}^+ = 0.020 \pm 0.006 \pm 0.001$}.
The measurements for the decay rates and \CP asymmetries are consistent with and more precise than the
corresponding analysis from the BaBar collaboration~\cite{Aubert:2009yw}.
In the future they will help to further constrain $\gamma$ and the related hadronic parameters for this system.

\subsection{$\gamma$ combination}

As stated above the best precision on $\gamma$ is obtained by combining information from many
$B\to DK$ decay modes. An updated combination from LHCb has recently been
performed~\cite{LHCb-CONF-2017-004} that utilises 85 observables and contains 37 parameters. It 
includes the $B^\pm \to D^{(*)0}K^{\pm}$ and $B^\pm \to DK^{\ast \pm}$ results described above\footnote{In fact,
only the $B^\pm \to DK^{\ast \pm}$ results from Ref.~\cite{LHCb-CONF-2016-014} are used.} 
and an updated decay-time-dependent
measurement of the \CP asymmetry in $\Bs\to\Ds\Kp$ decays~\cite{LHCb-CONF-2016-015}. The
final measurement is $\gamma = {76.8^{+5.1}_{-5.7}}^\circ$. The HFLAV collaboration have
combined this with existing measurements from the B-factories to obtain
\mbox{$\gamma = {76.2^{+4.7}_{-5.0}}^\circ$}, where the precision is
dominated by the LHCb measurement. Many more updates of $B\to DK$ channels can be expected with Run 2.
The aim is to have sub-degree-level precision at the end of the LHCb phase 1 upgrade in 2024 by which point LHCb will have collected 50\invfb.

\begin{figure}[t]
\centering
\includegraphics[width=0.45\linewidth]{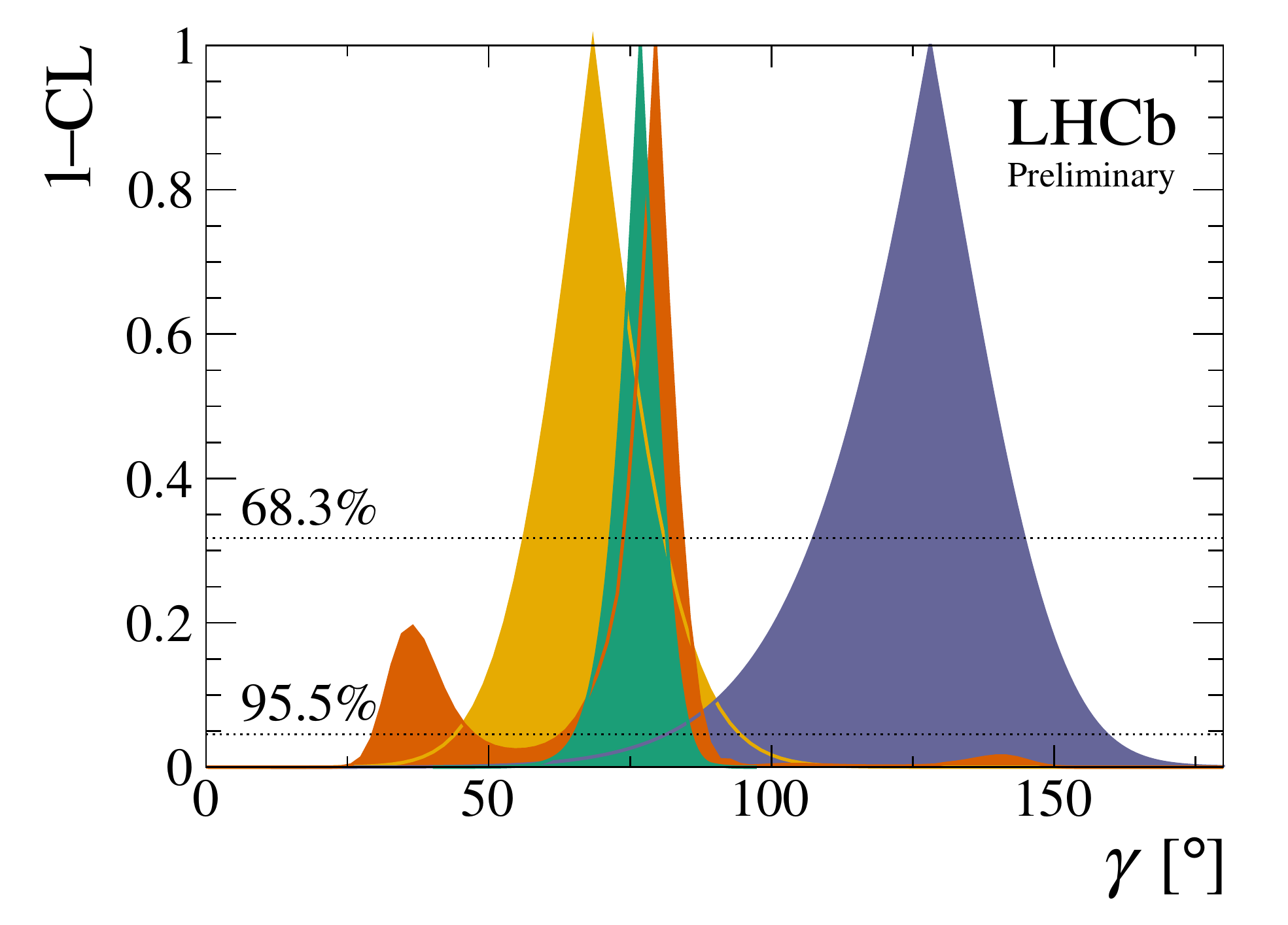}
\caption{\label{fig:comb}\small  1-CL plot, using the profile likelihood method, for the $\gamma$ combinations split
by analysis method.  (yellow) GGSZ methods, (orange) GLW/ADS methods, (blue) other methods and (green) the full combination.}
\end{figure}

\subsection{First evidence of \CP violation in the baryon system}

\begin{figure}[t]%
\begin{center}
  \parbox{2.1in}{\includegraphics[width=\linewidth]{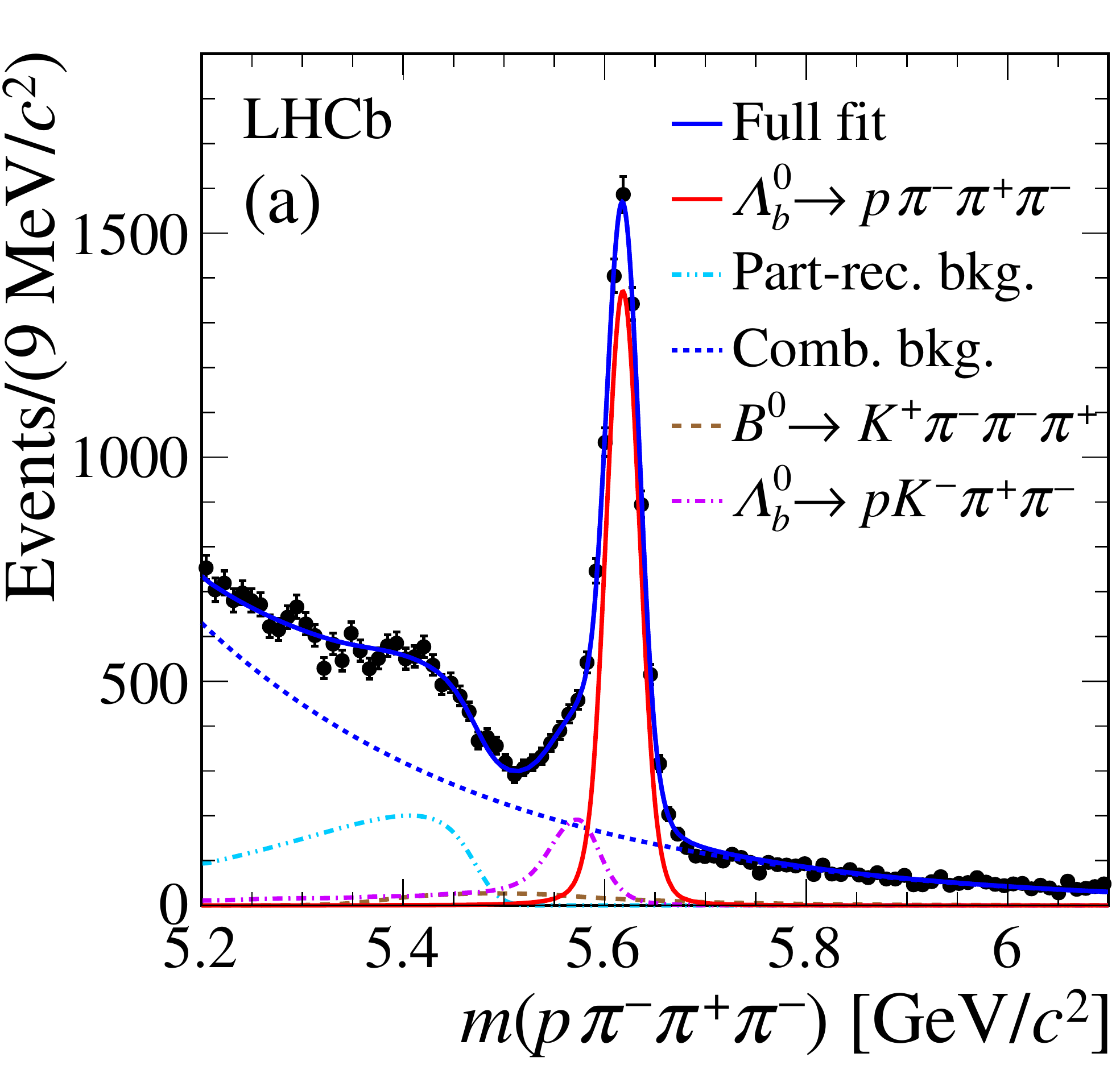}\figsubcap{a}}
  \hspace*{4pt}
  \parbox{2.1in}{\includegraphics[width=\linewidth]{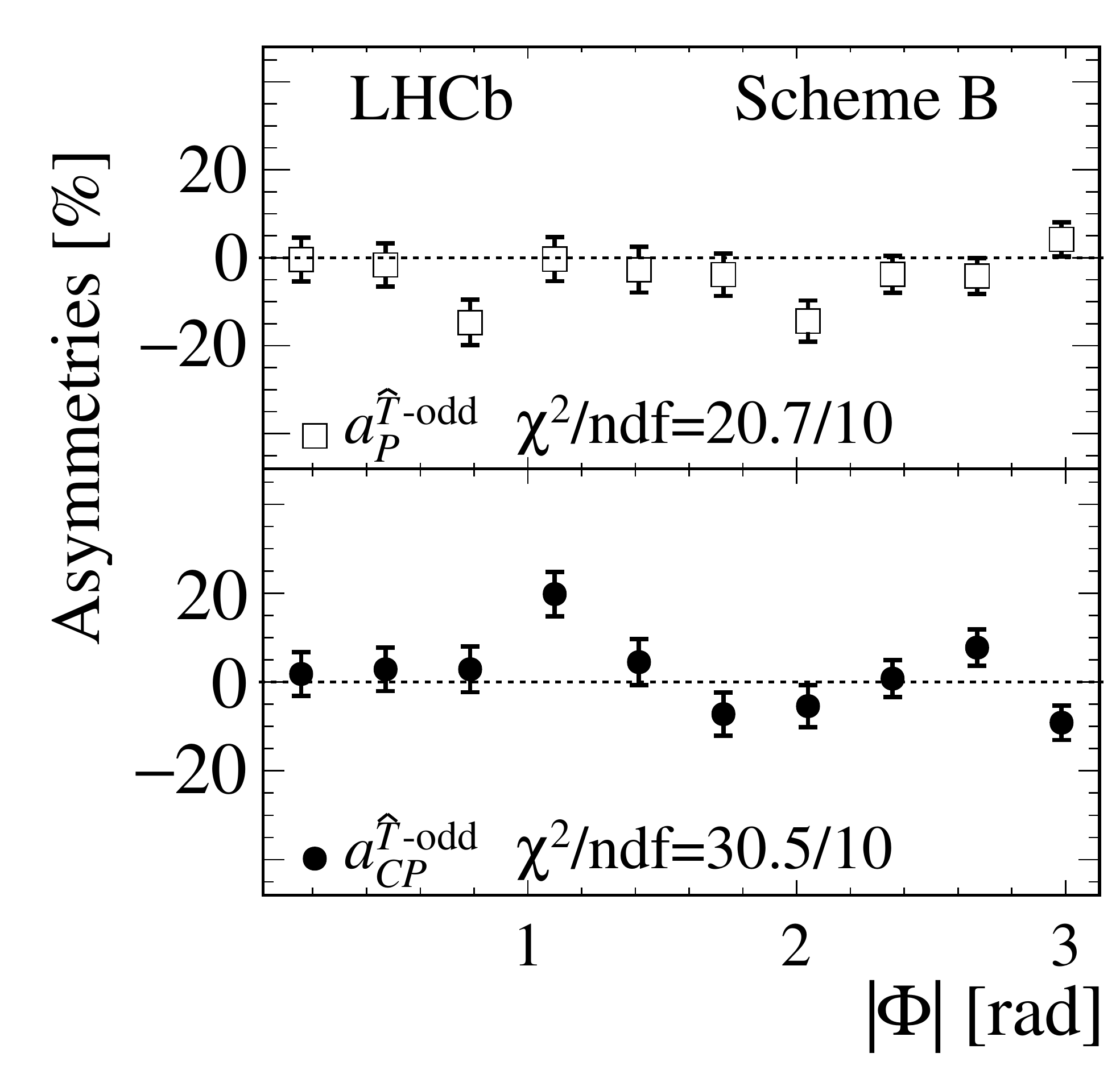}\figsubcap{b}}
\caption{\label{fig:Lb}\small  (a) The invariant mass distributions for $\Lb\to p\pi\pi\pi$ used to extract the 
asymmetries. (b) The $P$-odd and \CP-odd asymmetries as a function of $|\Phi|$, the magnitude of
angle between the 
$p\pi^-_{\rm fast}$ and $\pi^-_{\rm slow}\pi^+$ decay planes in the \Lb rest frame.}
\end{center}
\end{figure}

So far there has been no observation of \CP violation in $b$ baryon decays. However,
since they are governed by
the same quark-level transitions as meson decays there is potential for non-zero effects
in the SM~\cite{Bensalem:2002pz,Gronau:2015gha,Bigi:2016jii}. For example, charmless decays of
$b$ baryons have  contributions of similar magnitude from both tree and penguin decays, which give rise
to sensitivity to the CKM angle $\alpha$.
No sign of \CP violation has been found in such charmless two~\cite{Aaltonen:2014vra} or
three-body~\cite{LHCb-PAPER-2013-061,LHCb-PAPER-2016-004} $\Lb$ or $\Xi_b$
baryon decays, however, a recent study~\cite{LHCb-PAPER-2016-030} of four-body
$\Lb \to p h^-h^+h^-$ decays has revealed the first evidence for \CP-violation in the baryon sector.

The measurement is performed by using the four-body decay topology to compute 
triple products, which are odd under  the motion reversal operator, $\hat{T}$.\footnote{The $\hat{T}$
operator is equivalent to the parity operation for spinless particles.}
The triple products are defined as $C_{\hat{T}} = \vec{p}_p\cdot(\vec{p}_{h_1^-}\times\vec{p}_{h_2^+})$
and $\overline{C}_{\hat{T}} = \vec{p}_{\overline{p}}\cdot(\vec{p}_{h_1^+}\times\vec{p}_{h_2^-})$ for $\Lb$
and $\overline{\Lambda}_b^0$ decays, respectively.
The corresponding $\hat{T}$-odd asymmetries are 
$A_{\hat{T}}(C_{\hat{T}}) = \frac{N(C_{\hat{T}} > 0) - N(C_{\hat{T}} < 0)}{N(C_{\hat{T}} > 0) + N(C_{\hat{T}} < 0)}$
and
$\overline{A}_{\hat{T}}(\overline{C}_{\hat{T}}) = \frac{N(-\overline{C}_{\hat{T}} > 0) - N(-\overline{C}_{\hat{T}} < 0)}{N(-\overline{C}_{\hat{T}} > 0) + N(-\overline{C}_{\hat{T}} < 0)}$, 
where $N$ refers to the number of observed $\Lb \to p h^-h^+h^-$ candidates.
From these $\hat{T}$-odd asymmetries is is possible to build $P$-odd and \CP-odd observables, 
defined as $a_{P}^{\hat{T}-{\rm odd}} = \frac{1}{2}(A_{\hat{T}} + \overline{A}_{\hat{T}})$ and 
$a_{\CP}^{\hat{T}-{\rm odd}} = \frac{1}{2}(A_{\hat{T}} - \overline{A}_{\hat{T}})$, respectively.
The \CP-odd observable is insensitive to production and detection asymmetries that affect standard \CP-asymmetries
(e.g., those based on event yields) and is also formed from a different combination of strong and weak phases.

Figure~\ref{fig:Lb}a shows the $\Lb\to p\pi\pi\pi$ invariant mass from Ref.~\cite{LHCb-PAPER-2016-030} that
has been used to measure $a_{\CP}^{\hat{T}-{\rm odd}}$.
The global measurement is consistent with \CP symmetry but it has been noted~\cite{Durieux:2015zwa}
that there is increased sensitivity to \CP violating effects by looking at differential distributions. 
Figure~\ref{fig:Lb}b shows $a_{\CP}^{\hat{T}-{\rm odd}}$ as a function of the phase space of the 
$\Lb\to p\pi\pi\pi$ decay. Using this and an alternative binning scheme,
the p-value for the \CP-symmetry hypothesis is evaluated as $9.8\times 10^{-4}$, which is 
equivalent to a $3.3\sigma$ deviation of $a_{\CP}^{\hat{T}-{\rm odd}}$ from zero. This constitutes the first evidence
of \CP violation in baryon decays.
With the larger data samples to be collected in Run 2 and beyond it will be possible to perform a
full amplitude analysis of the $\Lb\to p\pi\pi\pi$ channel to understand where the
\CP asymmetry arises in the phase space of the decay.

Similar methods using $\hat{T}$-odd observables have been used to search for \CP violation in rare $\Lb$ decays~\cite{LHCb-PAPER-2016-002,LHCb-PAPER-2016-059}
and the charm system~\cite{LHCb-PAPER-2014-046}. In each
case the results are consistent with the hypothesis of \CP conservation.

\section{Study of $b$-baryon oscillations}

As noted in Section 1 the origin of the baryon asymmetry in the Universe is unclear. 
Baryon number violation (BNV) has never been seen experimentally, with strong constraints
imposed by the measured proton and bound-neutron lifetimes.
However, beyond-the-SM models containing flavour-diagonal six-fermion
vertices~\cite{Smith:2011rp,Durieux:2012gj,McKeen:2015cuz,Aitken:2017wie} could permit BNV without
violating existing constraints.
Unambiguous experimental observation of such BNV would be the observation of baryon-antibaryon
oscillations of hadrons containing quarks of all three generations $(i.e., usb)$, such as the $\Xi_b^0$ baryon.

\begin{figure}[t]
\begin{center}
\includegraphics[width=0.6\linewidth]{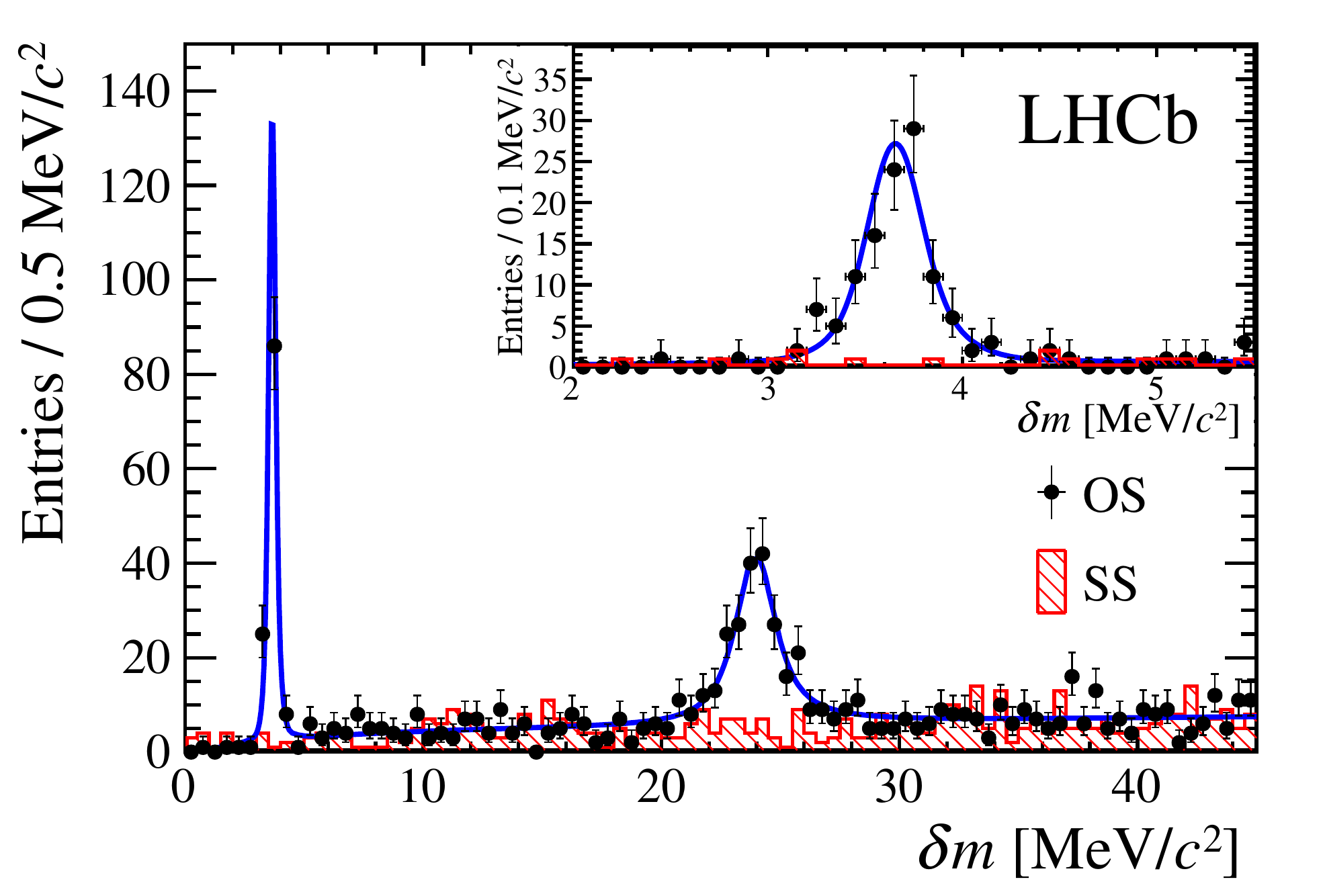}
\end{center}
\caption{\label{fig:Xib}\small Spectra of the mass difference $\delta m = m(\Xi_b^0\pi) - m(\Xi^0_b)) - m_\pi$
for the OS-tagged  (black points with error bars) and SS-tagged
 (red, hatched histogram) decays. Inset: detail of the region $2.0 < \delta m < 5.5$ \mevcc.}
\end{figure}

A new result~\cite{LHCb-PAPER-2017-023} from the LHCb collaboration measures the 
decay-time-dependent ratio between the rates of same-sign (SS) and opposite-sign (OS) decays of the
$\Xi_b^0$ baryon. The ratio is defined as $R(t) = \frac{\Gamma(\Xi_b^0\to\overline{\Xi}_c^-\pi^+)}{\Gamma(\Xi_b^0\to\Xi_c^+\pi^-)} \approx (\omega t)^2$, where $\omega$ is the mixing frequency.
Here, SS (OS) means that the charge of the proton from the charged $\Xi_c\to pK\pi$  decay is the same
(opposite) to the charge of the pion from the strong decay of the excited $\Xi_b^{\prime,*}$ baryons
that tags the initial flavour of the $\Xi_b^0$ candidate. Figure~\ref{fig:Xib} shows the mass difference
distribution for the LHCb data. The two narrow peaks visible near threshold in the OS-tag sample
are due to the excited $\Xi_b^{\prime,*}$ baryons.
The red histogram shows the corresponding distribution for the SS tags, which is consistent with the
background-only hypothesis.
In seven bins of decay time the ratio $R(t)$ is evaluated using a likelihood fit to the mass distribution,
which allows an upper limit to be set on the mixing frequency. The limit is
 $\omega < 0.08$ ps$^{-1}$ at 95\% CL, determined using a likelihood ratio test and the CLs method.
 %~\cite{LHCB-PAPER-2014-061}

\section{Charm}

Studies of the charm system provide the only way to investigate \CP violation using up-type quarks.
As in the neutral \B meson system, neutral \D mesons undergo oscillations, but in this case they are
dominated by long-distance effects and the mixing parameters
($x=\Delta m/\Gamma$, $y = \Delta\Gamma/(2\Gamma)$) are expected to be small. Likewise, very small
\CP-violating effects are expected~\cite{Bhattacharya:2012ah}. As yet there is no evidence for 
\CP violation in the charm system, but the very large data samples being collected by the LHCb 
experiment provide an ideal place to precisely test theoretical predictions and search for new sources of \CP
violation.

At hadron colliders it is possible to tag the initial flavour of $D^0$ mesons using two different methods,
depending on how they were produced. The flavour is determined either by 
the charge of the pion from promptly-produced \mbox{$D^{*+} \to D^0\pi^+$} decays or the charge of the muon from
semileptonic \B decays ($B \to D^0\mu^-X$). These independent samples have different properties in terms of 
background and reconstruction efficiencies, leading to different systematic uncertainties in the final measurements.

The large yields available in the charm system can also be used to search for \CP violation
in strong interactions. The LHCb collaboration has recently used $D^+_{(s)}\to\pi^+\pi^+\pi^-$
decays to search for \CP violating $\eta^{(\prime)}\to\pi^+\pi^-$ decays~\cite{LHCb-PAPER-2016-046},
finding no signal in the $\pi^+\pi^-$ mass spectrum. This constrains the branching
fractions to be less than $\sim10^{-5}$ at 90\% CL,
which are comparable with or better than existing limits~\cite{Patrignani:2016xqp}.

\subsection{Direct \CP violation in the charm system}
\label{sec:charm_direct}

Direct \CP violation in the charm system can be explored by measuring the asymmetries
in the decay of \Dz mesons to \CP eigenstates (i.e., $\Dz\to h^+ h^-$).
Experimentally the raw yield asymmetry is measured,
\mbox{$A_{\rm raw} \equiv \frac{N(\Dz\to h^+ h^-) - N(\Dzb\to h^+ h^-)}{N(\Dz\to h^+ h^-) + N(\Dzb\to h^+ h^-)}$},
and subsequently corrected for small production and detection asymmetries using data 
control modes in order to determine the \CP asymmetry, 
\mbox{$A_{\CP}(\Dz\to h^+ h^-) = A_{\rm raw}(\Dz\to h^+ h^-) - A_{\rm P}(D^{*+}) - A_{\rm D}(\pi_s^+)$}.

\begin{figure}[t]%
\begin{center}
  \parbox{2.1in}{\includegraphics[width=\linewidth]{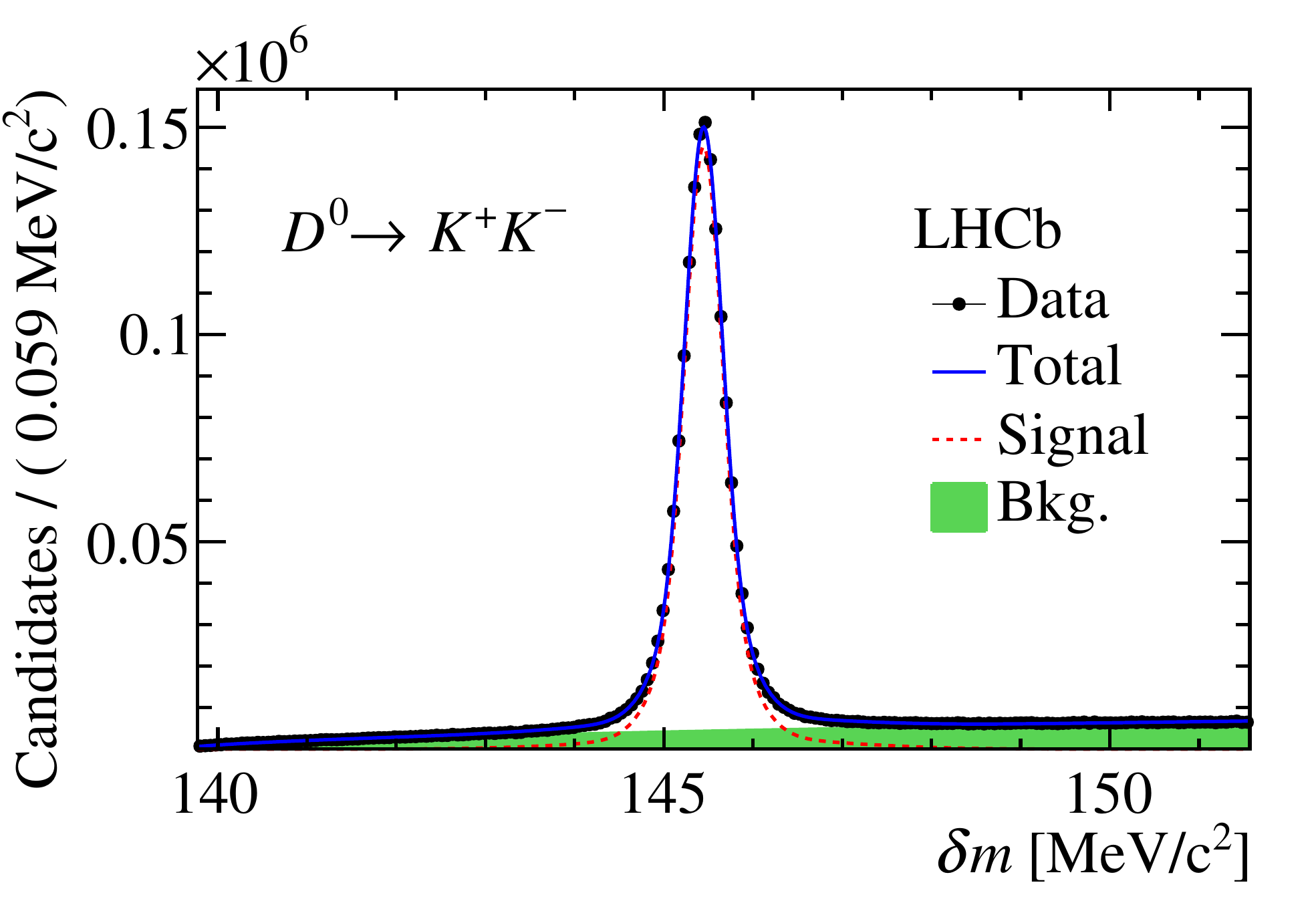}\figsubcap{a}}
  \hspace*{4pt}
  \parbox{2.1in}{\includegraphics[width=0.9\linewidth]{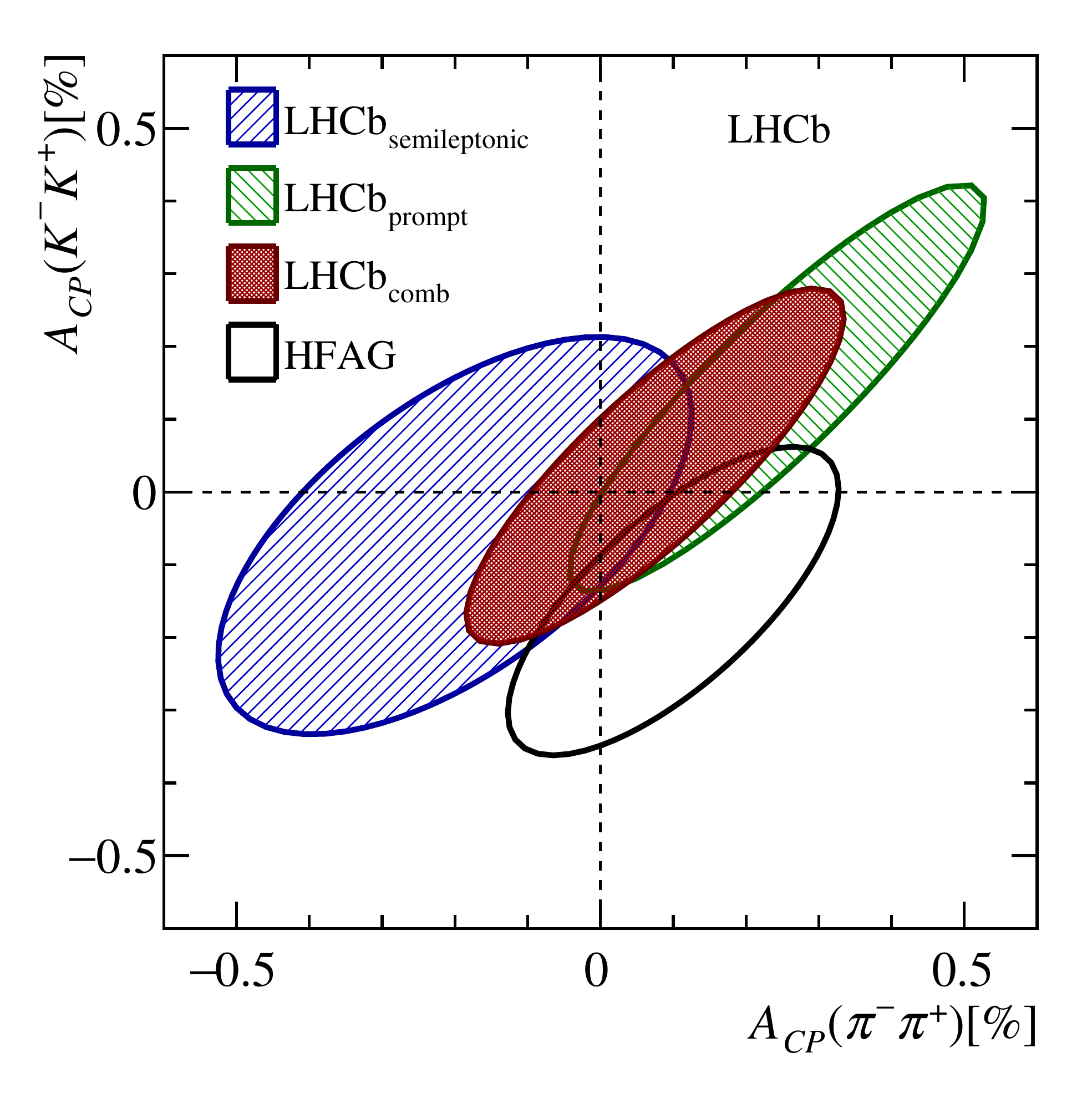}\figsubcap{b}}
\caption{\label{fig:charm_direct}\small  (a) Distribution of the mass difference between the $D^*$ and $D^0$ 
mesons in the prompt sample. (b) \CP violating asymmetries for $D^0\to\Kp\Km$ and $D^0\to\pi^+\pi^-$
decays using the prompt and semileptonic-tagged samples. The 68\% confidence level contours are displayed where the statistical and systematic uncertainties are added in quadrature.}
\end{center}
\end{figure}

The LHCb collaboration has recently measured
$A_{\CP}(\Dz\to K^+ K^-)$ using a prompt-tagged sample
(Figure~\ref{fig:charm_direct}a)~\cite{LHCb-PAPER-2016-035}.
While still statistically limited, the dominant systematic uncertainty in the result relates to control
of the nuisance asymmetries.
Combining $A_{\CP}(\Dz\to K^+ K^-)$ with an earlier measurement
of  the difference in \CP asymmetries between the kaon and pion modes, 
\mbox{$\Delta A_{\CP} = A_{\CP}(\Dz\to K^+ K^-) - A_{\CP}(\Dz\to \pi^+\pi^-)$}, allows $A_{\CP}(\Dz\to \pi^+\pi^-)$ to be extracted. Figure~\ref{fig:charm_direct}b
shows the result for the two \CP asymmetries, which are consistent with \CP symmetry. These results
have been combined with independent measurements from the semileptonic-tagged
sample~\cite{LHCb-PAPER-2014-013}
to provide an overall LHCb result of $A_{\CP}(\Dz\to K^+ K^-) = (0.04 \pm 0.12 \pm 0.10)\%$
and \mbox{$A_{\CP}(\Dz\to K^+ K^-) = (0.07 \pm 0.14 \pm 0.11)\%$}, where the first uncertainty is statistical 
and the second systematic. These results are now approaching the per-mille level of uncertainty
but still do not show any signs of \CP violation.
Likewise, there are no indications of \CP violation in other modes such as $\D^\pm\to\eta^\prime\pi^\pm$
and $\D_s^\pm\to\eta^\prime\pi^\pm$~\cite{LHCb-PAPER-2016-041}.
A search for \CP violation in $\D^0\to\pi^+\pi^-\pi^+\pi^-$ decays using an energy test~\cite{LHCb-PAPER-2016-044}
reports a mild tension with \CP-symmetry that requires further exploration with Run 2 data.

\subsection{Indirect \CP violation}

As discussed in above, the mixing parameters and decay-time-integrated \CP asymmetries in the 
charm system are small. Based upon this, a measurement of the decay-time-dependent
\CP asymmetry of \Dz decays to \CP eigenstates is sensitive to the indirect \CP violating parameter
$A_{\Gamma}$. Here, $A_{\Gamma} \equiv \frac{\hat{\Gamma}_{\Dz\to f} - \hat{\Gamma}_{\Dzb\to f}}{\hat{\Gamma}_{\Dz\to f} + \hat{\Gamma}_{\Dzb\to f}}$, is the asymmetry between the inverse effective lifetimes of the \Dz 
meson and its antiparticle. Within the SM, its magnitude is expected to be $\lesssim 5\times10^{-3}$
and independent of the final state, $f$, owing to the small \CP violation in decay.

\begin{figure}[t]%
\begin{center}
  \parbox{2.1in}{\includegraphics[width=1.1\linewidth]{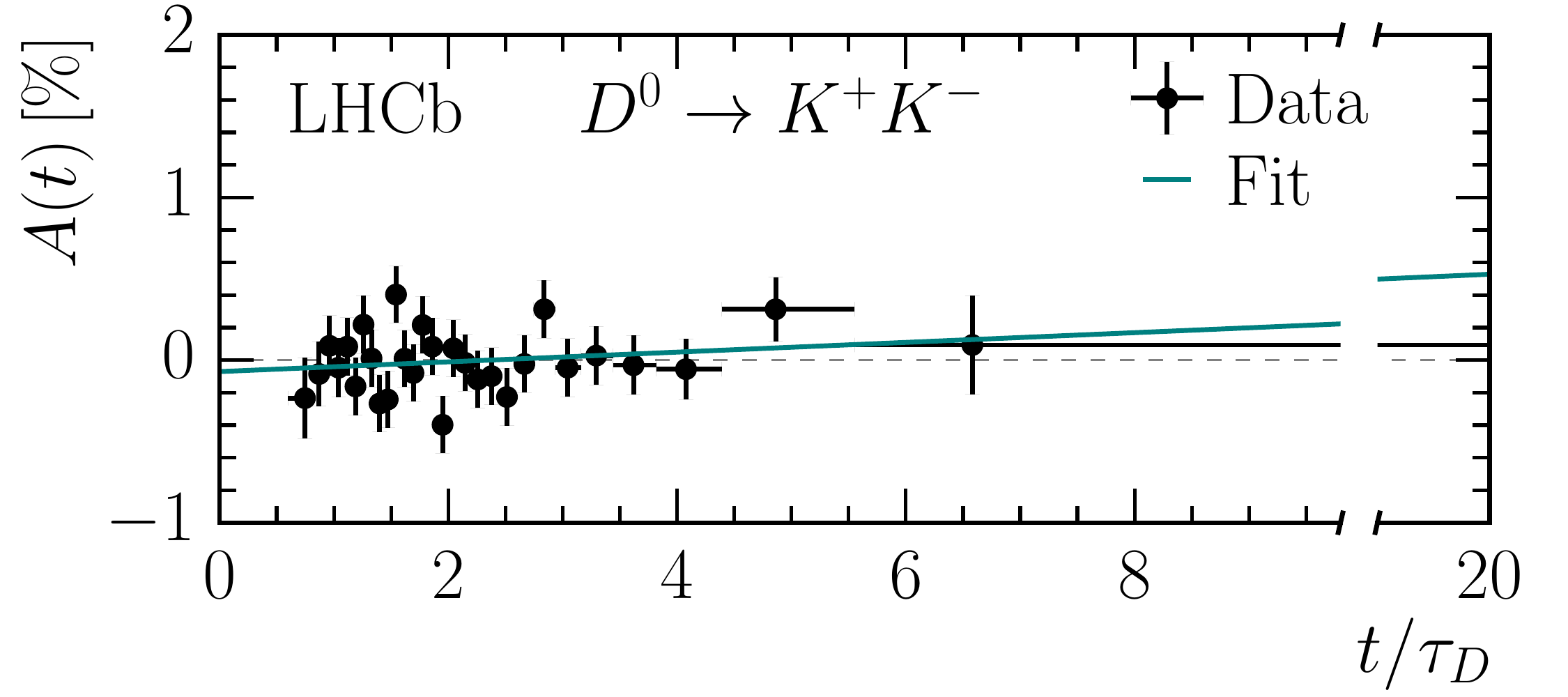}\figsubcap{a}}
  \hspace*{4pt}
  \parbox{2.1in}{\includegraphics[width=1.1\linewidth]{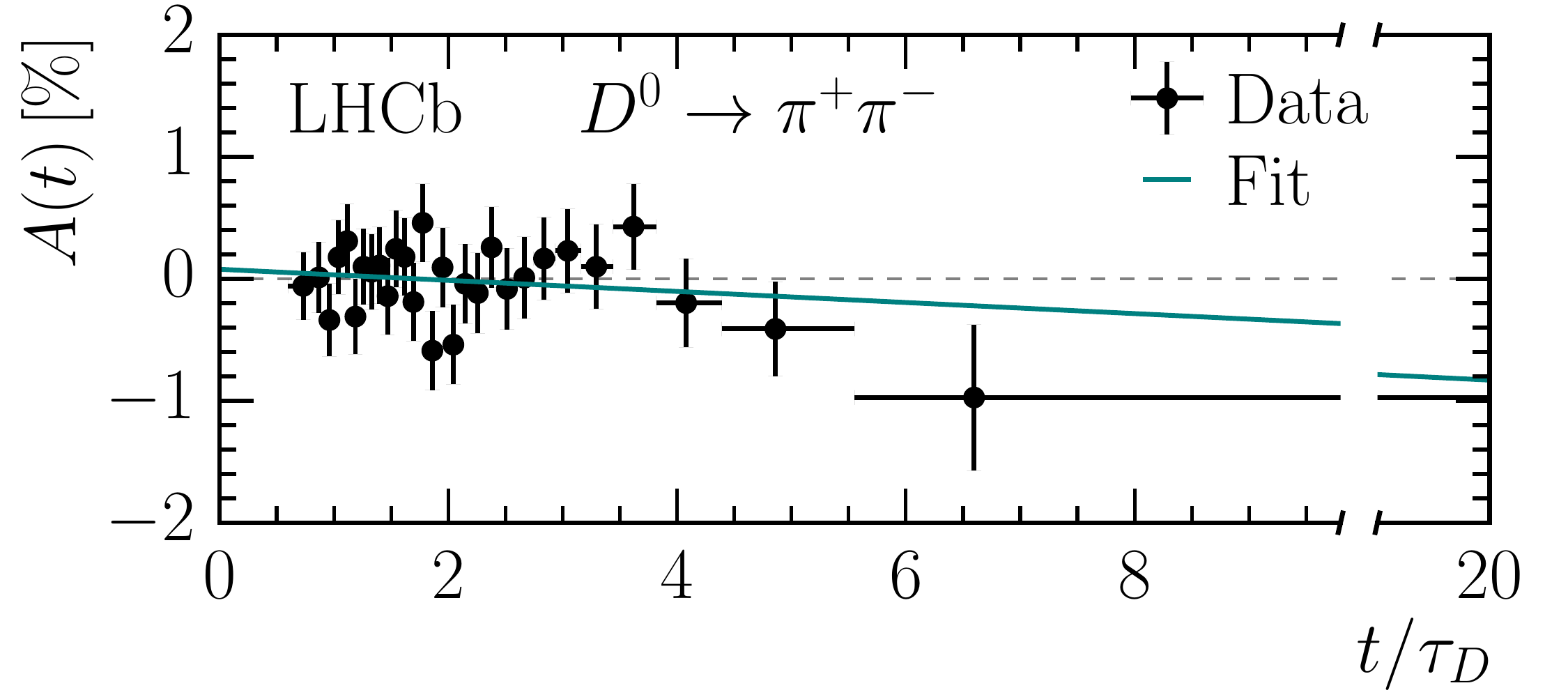}\figsubcap{b}}
\caption{\label{fig:charm_indirect}\small  Measured asymmetry $A(t)$ in bins of $t/\tau_D$, where $\tau_D$ is the 
world average value for the $\D^0$ lifetime~\cite{Amhis:2016xyh},
for (a) $D^0 \to \Kp\Km$ and (b) $D^0 \to \pi^+\pi^-$ decays.
The solid line shows a linear fit to the data, with a slope equal to the best estimate of $-A_\Gamma$.}
\end{center}
\end{figure}

The LHCb collaboration has recently measured $A_\Gamma$ using a prompt-tagged sample
of $\Dz\to\Kp\Km$ and $\Dz\to\pi^+\pi^-$ decays~\cite{LHCb-PAPER-2016-063}. Two 
different methods are used to measure the asymmetry, an approach that is binned in the \Dz decay time and
an unbinned method. Both approaches use the large sample of $D\to K\pi$ decays to control production and
detection asymmetries. They each give consistent measurements, with the binned result being
slightly more precise and chosen as the nominal result.
Figure~\ref{fig:charm_indirect} shows the binned asymmetries as a function of decay time for the kaon and pion 
decay modes. The value of $A_\Gamma$ is given by the gradient of the linear fit to these asymmetries.
These values are combined to give $A_\Gamma = (-0.13 \pm 0.28 \pm 0.10)\times 10^{-3}$.
They have subsequently been combined with the semileptonic-tagged
sample~\cite{LHCb-PAPER-2015-030} to obtain $A_\Gamma = (-0.29 \pm 0.28)\times 10^{-3}$.
This is the most precise measurement of a 
\CP-violating observable in the charm system ever made.

\subsection{Charm mixing and indirect \CP violation}
\label{sec:charm_direct}

\begin{figure}[t]%
\begin{center}
  \parbox{1.55in}{\includegraphics[width=\linewidth]{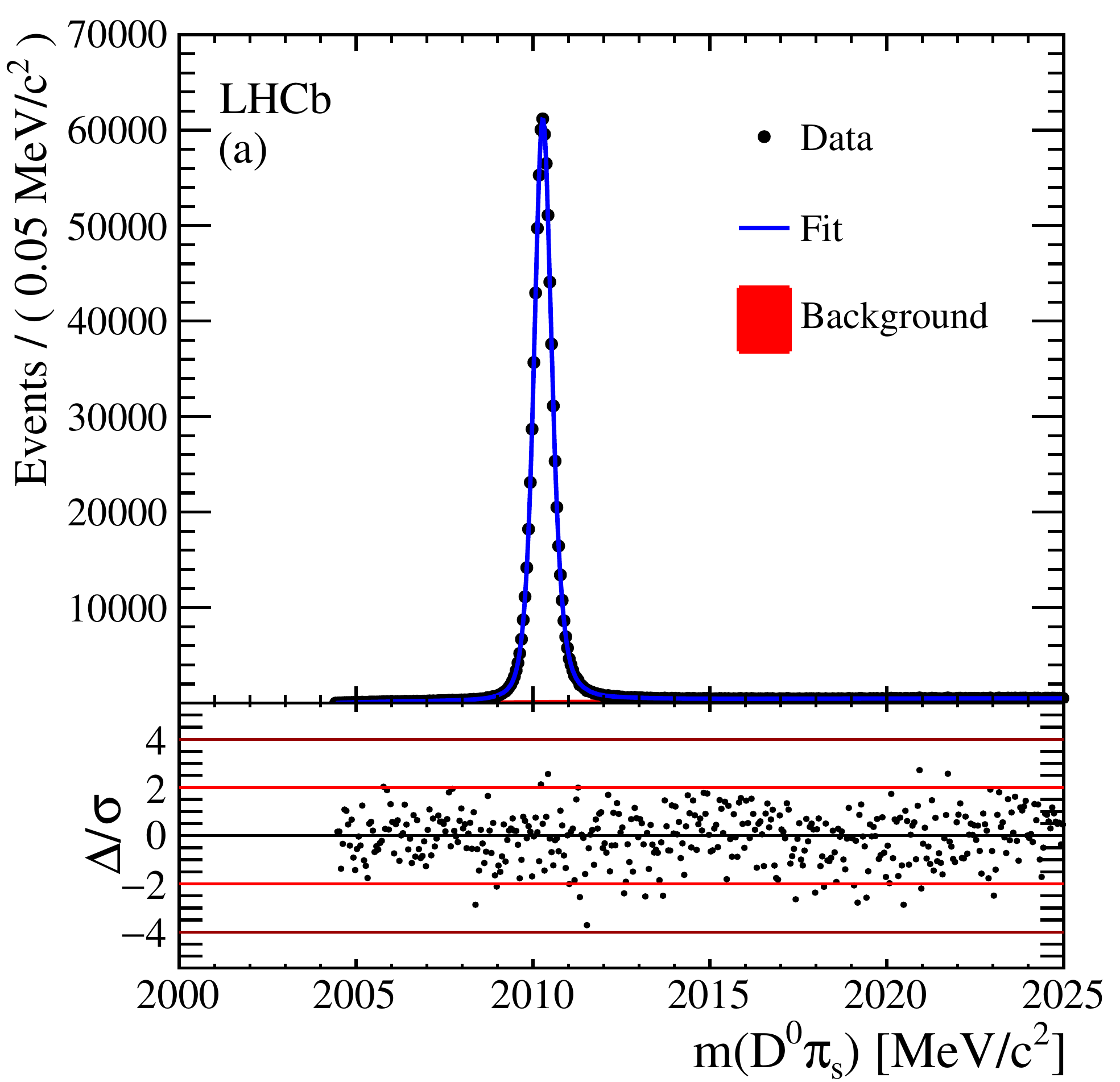}\figsubcap{a}}
  \hspace*{4pt}
  \parbox{1.55in}{\includegraphics[width=\linewidth]{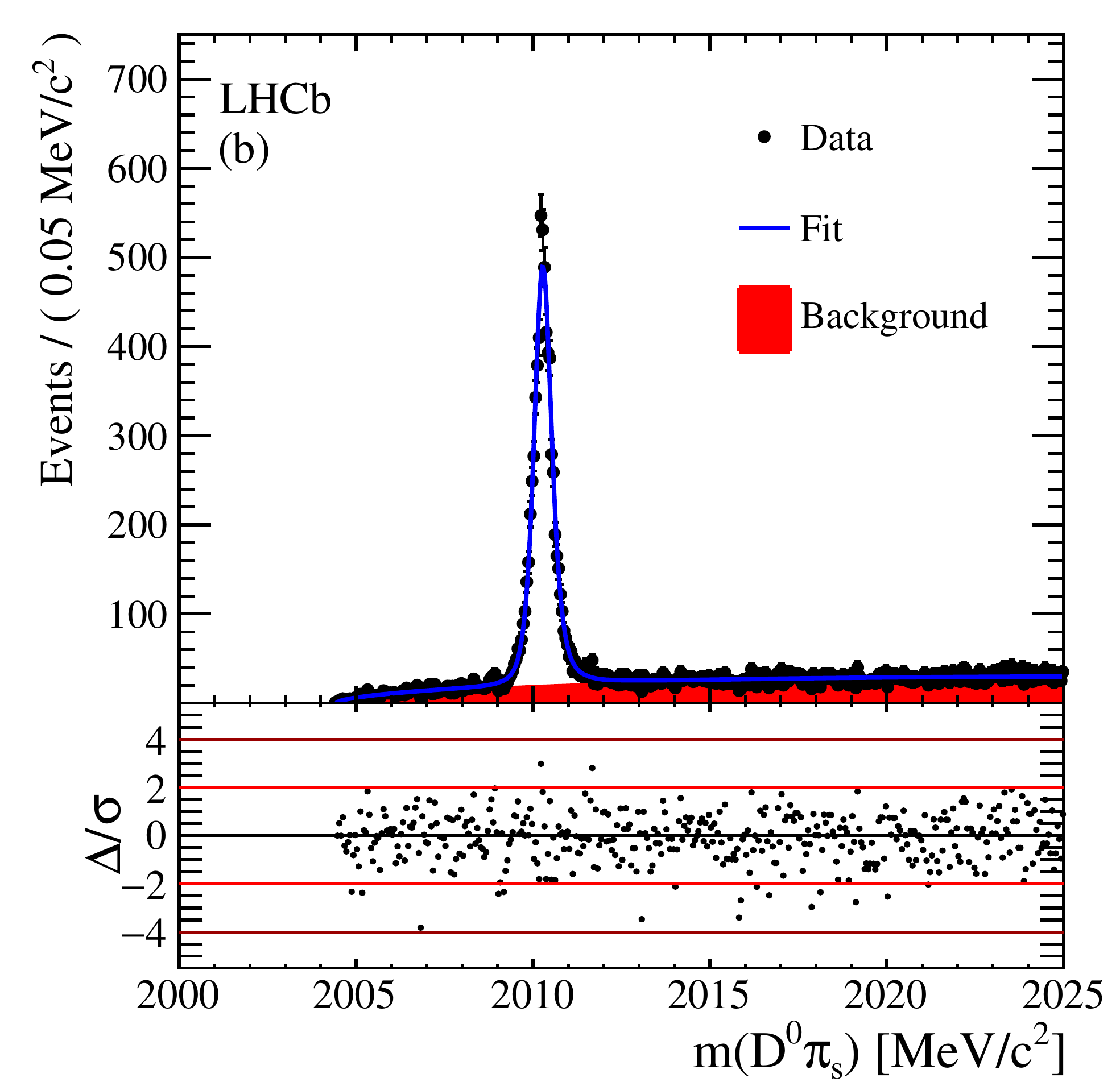}\figsubcap{b}}
  \hspace*{4pt}
  \parbox{1.55in}{\includegraphics[width=\linewidth]{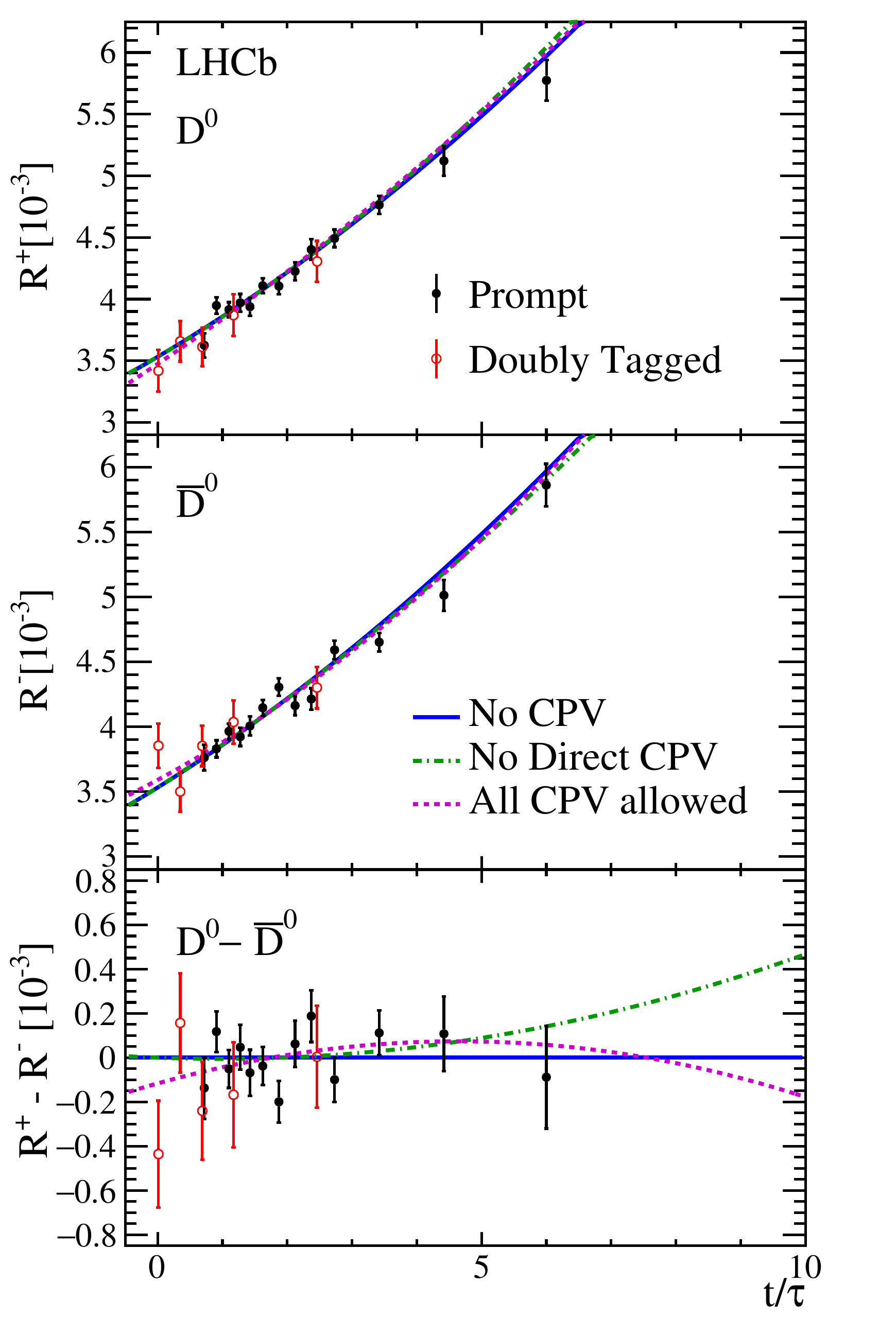}\figsubcap{c}}
\caption{\label{fig:charm_mixing}\small  Distribution of the $D^{*+}$ invariant mass 
in the double-tagged (DT) sample for (a) RS and (b) WS decays. (c) Efficiency corrected ratios of
WS/RS decays and fit projections for the DT  (red open circles) and prompt (black filled circles) samples.}
\end{center}
\end{figure}

The measurement of charm mixing parameters and a search for \CP violation can be performed by studying the 
decay-time-dependent ratio of yields of Cabibbo-suppressed to Cabibbo-favoured $D^0\to K\pi$
decays. This ratio is defined separately for \Dz and \Dzb decays and is related to the mixing parameters via
a second-order expansion of the mixing equations (assuming small mixing) given by
\mbox{$R(t)^\pm = R_D^\pm + \sqrt{R_D^\pm}y^\prime \frac{t}{\tau} + \frac{(x^{\prime\pm})^2 + (y^{\prime\pm})^2}{4}\left(\frac{t}{\tau}\right)^2$},
where $\tau$ is the average \Dz lifetime and the $x^\prime, y^\prime$ have been rotated from the 
nominal mixing parameters by the strong phase in the $\D\to K\pi$ decay.
A recent result~\cite{LHCb-PAPER-2016-033} from the LHCb collaboration uses a
double-tagged (DT) technique (using the charge of the
pion and the muon from the semileptonic \B decays ($\Bd \to (D^{*+}\to D^0\pi^+)\mu^-X$)) to provide 
a very pure sample of events (Figure~\ref{fig:charm_mixing}a and b) that
cover a complementary region of the decay time spectrum than
has been previously studied using a prompt-sample~\cite{LHCb-PAPER-2013-053}.
Figure~\ref{fig:charm_mixing}c shows the ratios for \Dz and \Dzb decays and their difference for both the
DT and prompt samples. The ratios increase as a function of time, consistent with charm mixing,
allowing $x^\prime, y^\prime$ to be measured for the DT sample alone and both samples combined.
The complementary coverage of the DT and prompt sample gives an improved precision, by 10--20\%, for
the charm mixing parameters from the combined fit even though the DT analysis is based on almost 40 times fewer
candidates than the prompt analysis.
The data are consistent with the hypothesis of \CP conservation (both for decay and interference
of mixing and decay).

\section{Looking to the future}

\begin{figure}[t]
\centering
\includegraphics[width=0.5\linewidth]{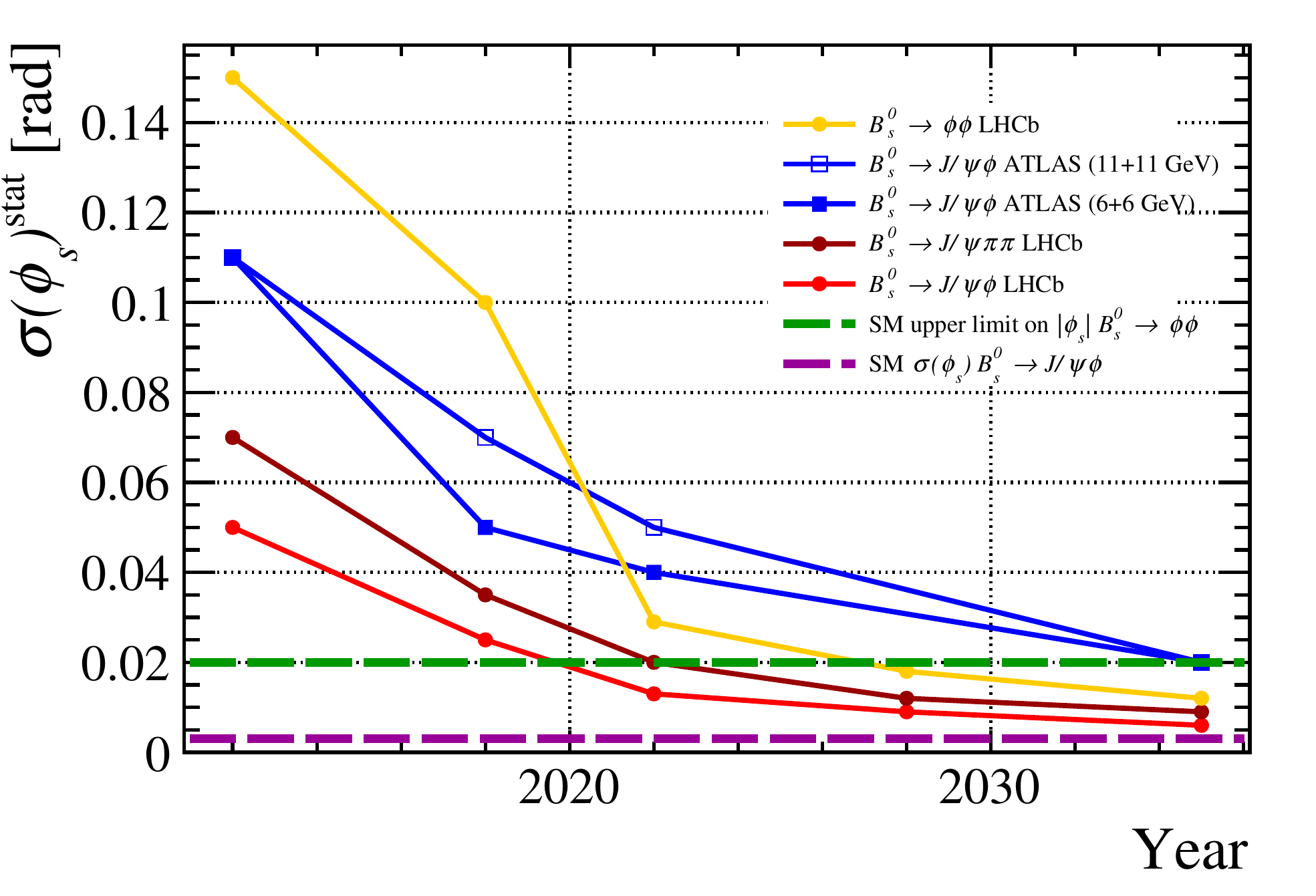}
\caption{\label{fig:projections}\small  Projection of how precision on \phis\ from LHCb
measurements will scale as a function of
time for different decay modes. Information taken from Ref.~\cite{LHCb-PUB-2014-040,ATL-PHYS-PUB-2013-010}.}
\end{figure}

The majority of beauty and charm sector measurements are
statistically limited, which strongly motivates the case for a new set of precision
measurements of \CP-violating observables to constrain the size of (or perhaps find) new physics
contributions.
The LHCb collaboration recently submitted an expression of interest~\cite{Aaij:2244311}
regarding future phase 1b and 2 upgrades that would operate during LHC runs 4 and 5, respectively.
It is expected that using the data collected with the improved detector it 
will be possible to improve the precision on $\gamma$ to 
0.4$^\circ$, $\phi_s$ to  9 mrad (see Figure~\ref{fig:projections})
and achieve precision on charm mixing and \CP
violation parameters at the level of $10^{-4}$.
These will allow unprecedented sensitivity to the small effects of new physics.
Crucially, as the precision improves it is essential to control hadronic effects
that can hide small  non-Standard Model effects.
Other studies can be found in Refs.~\cite{LHCb-PUB-2014-040,ATL-PHYS-PUB-2013-010}.

\section{Summary}
This year is the 40th anniversary of the discovery of the $b$ quark~\cite{Herb:1977ek}.
In that time there has been huge progress in using the physics of heavy-flavour hadrons to
make detailed studies of \CP-violation, both experimentally and theoretically.
So far all measurements are consistent with Standard Model predictions, confirming validity of
the  CKM mechanism. The LHCb experiment is now leading the way in terms of precision
measurements in the $b$ and $c$ sectors, with new explorations of \CP violation in baryons
just beginning. It will be fascinating to observe how the SM is able to withstand the next round
of precision measurements that will be made over the next decade.

\section*{Acknowledgements}

The author thanks the organisers of the Lepton Photon conference and acknowledges the support of
the Science and Technology Facilities Council (UK) grant ST/K004646/1.

%\clearpage
%\addcontentsline{toc}{section}{References}
%\setboolean{inbibliography}{true}
\bibliographystyle{ws-procs961x669}
%\bibliographystyle{ieeetr}
%\bibliography{LHCb-PAPER,LHCb-CONF,LHCb-DP,LHCb-TDR}
\bibliography{LHCb-PAPER-new,LHCb-CONF-new}

\end{document}